\documentclass[10pt,newlfont,aps,pra,superscriptaddress,twocolumn,showkeys]{revtex4}
\def\ea{{\it et al.}}
\def\hs{\hspace{0.5cm}}

\def\ft{\hspace{0.1cm}}

\def\Gkdd{${\cal G}_{1dd}$}

\def\Dkin{${\cal D}_{kin}$}
\def\Dkino{${\cal D}^{(0)}_{kin}$}
\def\aho{$a_{ho}$}
\def\GkD{${\cal G}_{1D}$}

\def\dGkdd{$\Delta {\cal G}_{1dd}$}
\def\hw{$\hbar\bar{\omega}$}
\usepackage{graphicx}
\usepackage{color}
\usepackage{amsbsy}
\usepackage{amsmath}
\begin{document}
\title{Characterization of the energy level-structure of a trapped dipolar Bose gas via mean-field parametric resonances}
\author{Asaad R. Sakhel}
\affiliation{Department of Physics, Faculty of Science, Balqa Applied University, Salt 19117, Jordan}
\author{Roger R. Sakhel}
\affiliation{Department of Physics, Faculty of Science, Isra University, Amman 11622, Jordan}
\begin{abstract}
We report parametric resonances (PRs) in the mean-field dynamics of a one-dimensional 
dipolar Bose-Einstein condensate (DBEC) in widely varying trapping geometries. The chief goal is 
to characterize the energy levels of this system by analytical methods and the significance of 
this study arises from the commonly known fact that in the presence of interactions the energy 
levels of a trapped BEC are hard to calculate analytically. The latter characterization is 
achieved by a matching of the PR energies to energy levels of the confining 
trap using perturbative methods. Further, this work reveals the role of the interplay between 
dipole-dipole interactions (DDI) and trapping geometry in defining the energies and amplitudes of 
the PRs. The PRs are induced by a negative Gaussian potential whose depth oscillates with time. 
Moreover, the DDI play a role in this induction. The dynamics of this system is modeled by the time-dependent Gross-
Pitaevskii equation (TDGPE) that is numerically solved by the Crank-Nicolson method. The PRs are 
discussed basing on analytical methods: first, it is shown that it is possible to reproduce 
PRs by the Lagrangian variational method that are similar to the ones obtained from 
TDGPE. Second, the energies at which the PRs arise are closely matched with the energy levels of 
the corresponding trap calculated by time-independent perturbation theory. Third, the most probable transitions 
between the trap energy levels yielding PRs are determined by time-dependent perturbation theory. 
The most significant result of this work is that we have been able to characterize the above 
mentioned energy levels of a DBEC in a complex trapping potential.
\end{abstract}
\date{\today}
\maketitle

\section{Introduction}

\hs The phenomenon of parametric resonances (PRs) is ubiquitious in nature and is a widely examined fundamental physical 
property. Today, PRs are one of the outstanding features observed in Bose-Einstein condensates (BECs) 
\cite{Balik:2009,Sakhel:2020,Nguyen:2019,Chen:2019,Zhu:2019,ZhengChun:2019,Sakhel:2018,
Molignini:2018,Robertson:2018,Lellouch:2017,Vidanovic:2012,Posazhennikova:2016,Kobyakov:2012,
Xue:2008,Engels:2007a,Nicolin:2007,Kraemer:2005,Tozzo:2005} where resonances are usually the result of modulating one of the system 
parameters such as the scattering length \cite{Chen:2018,Sabari:2018,Cairncross:2014,Vidanovic:2012,Vidanovic:2011}. 
They are also generated by external means such as a time-dependent trapping geometry \cite{Vidanovic:2012}, 
laser stirring \cite{Sakhel:2020}, and laser-intensity modulation \cite{Balik:2009}. 
Moreover, PRs have been shown to occur in classical systems \cite{Arnold:1989,Landau:1976} 
as well as quantum devices such as birefringent optical fibers \cite{Armaroli:2013}, 
magnetometers \cite{Beato:2020}, superconducting wave guides \cite{Fomin:1997}, and quantum dots \cite{Calvo:1999,Hackenbroich:1998}. 
Their importance has also been demonstrated in transatlantic telecommunication fiber optics \cite{Matera:1993} 
in connection to a modulational instability.

\hs In this work, we explore PRs in a dipolar BEC (DBEC) within a setting of large-sized traps 
that would allow an examination of the effects of long-range interactions. It is known that
dipole-dipole interactions (DDI) are capable of shifting the frequency of PRs 
\cite{Bismut:2010,Goral:2002,Balik:2009}. This shift is sensitive to the trapping geometry 
\cite{Bismut:2010}. As a result, one concludes that the interplay between trapping geometry 
and DDI determines the energies at which PRs occur. These energies are equivalent to the 
energies of the trap levels; the trap in which the DBEC is confined. Within this context then,
the chief goal is to characterize the energy-level structure (ELS) of a DBEC in a trap of large 
size by matching the PR energies to energy levels of this trap using perturbative methods. Herein, 
the importance of PRs is revealed in characterizing the shape of the BEC trap and its energy 
levels \cite{Balik:2009}. The significance of the present investigation arises from the fact 
that it facilitates an evaluation of the above-named structure in complex potentials. This is
because it is known that, in the presence of interactions, it is rather hard to analytically 
calculate the energy levels of a trapped BEC. Moreover, via the present research, we hope to 
motivate future experiments that could determine the ELS of more complicated traps that may 
be engineered in the future. That said, the usefulness of the present examination is revealed 
from the latter statements. Another goal is to reveal the interplay between trapping geometry and DDI 
(cf. \cite{Mishra:2016,Sabari:2018,Waechtler:2016,Schulz:2015}) in defining the frequencies at 
which PRs occur. Effects of trapping geometry have already been studied earlier such as the 
influence of the trap aspect ratio on the oscillation frequencies \cite{Lima:2011} and stability 
of a DBEC \cite{Ancilotto:2014,Wilson:2011}, except that it hasn't been related to the structure 
of the trap energy levels.

\hs At present, we consider a one-dimensional (1D) trapped DBEC that is driven by a negative 
Gaussian potential (NGP) whose depth is periodically modulated with time. The latter system is 
simulated numerically in different traps using the mean-field time-dependent Gross-Pitaveskii 
equation (TDGPE). The 1D DBEC is scanned over a long range of DDI strengths in an attempt to detect 
PRs and to study the above mentioned interplay. In passing, it should be noted that the 1D Bose 
gas is a general, important, and well known system that can refer to many physical systems, such 
as optical fibers \cite{Armaroli:2013,Agrawal:1987,Abdullaev:1996,Ambomo:2008}. It is studied to 
reveal the physics in 1D which is strikingly different than in higher dimensions. 
The addition of an NGP is for the purpose of causing the particles to condense into lower energy 
levels, so to speak to ``catch particles", and then throw (excite) them to higher energy levels by 
the oscillating NGP depth. The oscillatory NGP has been earlier shown to work like a modulated 
contact interaction \cite{Sakhel:2018} and could therefore be viewed analogous to it. It should also be noted 
that an NGP has been used to induce a BEC and to study its growth dynamics \cite{Garrett:2011}. With the above NGP, 
a laser-light source is modelled whose intensity oscillates with time \cite{Clark:2015,Balik:2009} and provides a 
softer stirring of the BEC in energy space than the stirring by time-dependent spatial modulations. 
This method has rarely been used or mentioned in the BEC literature, and here it is demonstrated 
that it leads to significant excitations.

\hs The PRs are measured by a quantity that resembles the time average of the square of the 
kinetic energy called ``signal energy" \cite{Oppenheim:2015}. The signal energy is a term borrowed 
from engineering topics for the processing of an oscillating electrical signal. 
The signal energy has also been found very effective in revealing PRs in one of 
our earlier publications \cite{Sakhel:2020}.

\hs One motivation for the present examination arises from the work of Balik \ea\ \cite{Balik:2009}
who applied a CO${}_2$-laser generated optical-dipole trap to confine a sample of ${}^{87}$Rb 
atoms. By a modulation of the laser intensity, the authors were able to excite PRs whose
frequencies were found to shift by a change of the laser's modulation depth. In this regard,
we make an analogy by explaining the frequency shift as resulting from a change in the depth 
of the effective time-averaged potential of the trapped DBEC. The latter depth is controlled
by the DDI, and by scanning the DBEC over a broad range of the latter it has been possible to
observe PRs at values of DDI corresponding to the resonance frequencies. This project has been
heavily computational as it required a large number of runs to locate the DDI regimes where
PRs occur.

\hs The most significant result of this work is that we have been able to characterize 
the energy levels of a DBEC in a complex trapping potential, i.e., a trap to which an NGP is 
added. Other key results are detailed as follows. 

\begin{itemize} 
\item[(i)] By a variation of the DDI strength we detect mean-field PRs 
in the dynamics of a DBEC induced by an NGP whose depth is periodically modulated with time. 
The PRs obtained are an inherent feature deeply embedded in the physics of the DBEC as they 
could also be generated by other tools such as the Lagrangian variational method (LVM) 
\cite{Muruganandam:2012,Sakhel:2017,Sakhel:2018,Perez-Garcia:1996,Perez-Garcia:1997,Al-Jibbouri:2013,Falco:2009,Wei:2013}.

\item [(ii)] The positions of the PRs are determined by the ELS of 
the trapping geometry. It is shown that the energies at which PRs occur can be 
characterized by second-order perturbation theory thus allowing us to determine 
the above mentioned structure.

\item [(iii)] It is shown that the depth of the time-averaged effective potential is reduced 
by an increase in the repulsive DDI strength as a result of which the PR frequencies are shifted 
to larger values in the case of a harmonic oscillator (HO)$+$NGP. However, in the case of a 
BOX$+$NGP, the PR frequencies drop instead with rising DDI strength. Thus the interplay between 
trapping geometry and DDI controls the values of the DDI strength at which PRs arise in the 
signal energy \cite{Oppenheim:2015}. Moreover, the amplitude of the PRs 
decline with increasing DDI because the depth of the time-averaged mean-field effective potential 
is reduced with it.

\item[(iv)] It is found that the occurrence of a PR requires two conditions: (1) its 
energy should closely match one of the trap levels and (2) the transition probability to 
this level should be quite large.

\end{itemize}

\hs The organization of this paper is as follows. In Sec.\ft\ref{sec:method} the method is presented. In 
Sec.\ft\ref{sec:results} the results are presented and discussed and in Sec.\ft\ref{sec:origins-of-PRs} the 
origins of the PRs are explained. Finally in Sec.\ft\ref{sec:summary-and-conclusions} we summarize and conclude. 
In Appendix \ref{sec:numerics} technical details for the simulations are outlined and in Appendix 
\ref{sec:V1Dkunits} a measurement unit is derived.

\section{Method}\label{sec:method}

\hs In this section, only the rudiments of the method are outlined. The reader is therefore 
referred to Ref.\cite{Kumar:2015} for more details on the method and the numerics involved.

\subsection{Basic Units}

\hs In the present work, lengths and energies are in units of the trap $a_{ho}\,=\,\sqrt{\hbar/(m\bar{\omega})}$
and \hw, respectively, where $\bar{\omega}\,=\,(\omega_x \omega_y \omega_z)^{1/3}$ is the geometric 
average of the trapping frequencies along the coordinate axes, and $m$ is the mass of the atom. 
It should be however noted, that the modes in the $y$- and $z$-directions are frozen since the present 
system is described by the TDGPE that is reduced from 3D to 1D by integrating out the transverse 
directions. Time is unitless where $t=\bar{\omega} \tau$, $\tau$ being the time in seconds. 

\subsection{Systems}

\hs The systems considered are 1D strongly repulsive DBECs that are confined
in a few different trapping geometries. The confining potentials are power-law 
traps that have the form 

\begin{equation}
V_{tr}(x)\,=\,\frac{1}{2}\left|\frac{x}{L_x}\right|^{p},
\label{eq:one-dimensional-external-power-law-trap}
\end{equation}

where $p$ is the trapping exponent, and $L_x$ a length scale that shapes $V_{tr}(x)$ so that 
there is some flatness in the neighborhood of $x=0$ when $p$ is much larger than 2. 
$V_{tr}(x)$ is in units of \hw\ whereas $x$ and $L_x$ are in \aho. It should 
be emphasized that power-law traps have been realized in the United Kingdom using spatial 
light modulation \cite{Bruce:2011b}.

\hs The DBECs are excited by an NGP whose depth oscillates with time. Experimentally, the 
NGP is generated by the application of a focusing red-laser beam
\cite{Hammes:2002,Garrett:2011,Tuchendler:2008,Stamper-Kurn:1998,Comparat:2006,Jacob:2011,
Gustavson:2001,Barrett:2001,Schulz:2007}. The NGP has been considered in theoretical 
investigations \cite{Proukakis:2006,Diener:2002,Aioi:2011,Uncu:2008,Weitenberg:2011,Carpentier:2008} 
as well. Its importance has been demonstrated in an architecture of quantum 
computation \cite{Weitenberg:2011}, the transportation of BECs \cite{Gustavson:2001}, and the 
extraction of atoms from a BEC \cite{Diener:2002,Carpentier:2008}. Solitons \cite{Parker:2003} 
in an NGP and properties of a BEC in a harmonic trap plus an eccentric NGP \cite{Uncu:2008} 
have also been examined. The red-detuned laser beam interacts with the BEC in such a way so as 
to introduce an NGP into it by phase-imprinting \cite{Pethick:2008}. 
Adding to this the requirement of an oscillating intensity, the NGP is modelled by

\begin{equation}
V_{DT}(x,t)\,=\,[A\,+\,\delta A\cos(\Omega t)]\exp(-\beta x^2),
\label{eq:NGP-with-osc-depth}
\end{equation}

where $\Omega=2\pi f$ is the driving frequency, $A$ the principal depth, $\delta A$ the modulation 
amplitude, and as usual $1/\sqrt{\beta}$ a measure of the NGP width. $A$ and $\delta A$ are in 
units of \hw, $\beta$ in $a_{ho}^{-2}$, and $\Omega$ is in units of $\bar{\omega}$. As already noted 
in the introduction, the analog of the present excitation method is the periodic modulation of a 
scattering length such as $a(t)=a_{bg}\,+\,\delta a \sin(\omega t)$ \cite{Chen:2018,Sabari:2018,Cairncross:2014}, 
where $\delta a$ is the modulation amplitude, and $a_{bg}$ an unperturbed background. The analog of 
$a_{bg}$ is the $A$ and of $\delta a$ the $\delta A$. In one of our earlier publications \cite{Sakhel:2018}, 
it has been verified that the effects on the BEC arising from the NGP with oscillating depth 
[Eq.(\ref{eq:NGP-with-osc-depth})] are indeed similar to those from $a(t)$ above if $a_{bg} < 0$. 

\hs The time-modulation of the NGP depth in (\ref{eq:NGP-with-osc-depth}), 
that is $\delta V_{DT}(x,t)=\delta A\cos(\Omega t)e^{-\beta x^2}$, generates an oscillating force 
along the length of the DBEC that is given by the potential gradient

\begin{equation}
\delta F_x=-\frac{\partial\delta V_{DT}(x,t)}{\partial x}=2\beta x\delta A\cos(\Omega t)e^{-\beta x^2}.
\label{eq:driving-force}
\end{equation}

This force transfers a momentum from the oscillating NGP to the BEC that is symmetric about
$x=0$ and reads 

\begin{equation}
\Delta p_x=\int_0^t \delta F_xdt=2\beta xe^{-\beta x^2}\frac{\delta A\sin(\Omega t)}{\Omega}.
\end{equation}

The latter is maximal at $x\,=\,\pm1/\sqrt{2\beta}$ and minimal at $x=0$ and the edges of the 
trap. 

\subsection{Dipolar Interactions and their Control} 

\hs A note on the manipulation of the DDI is in order here. Consider the DDI 
potential given by \cite{Lahaye:2009,Kumar:2015}

\begin{equation}
U_{dd}(R)\,=\,C_{dd}\frac{(1-3\cos^2\theta)}{|\mathbf{R}|^3},
\end{equation}

where $\mathbf{R}=\mathbf{r}-\mathbf{r}^{\prime}$ is the relative position vector of two dipoles 
at $\mathbf{r}$ and $\mathbf{r}^{\prime}$, $\theta$ the angle between $\mathbf{R}$ and the orientation 
of the dipoles, $C_{dd}\,=\,E^2\alpha^2/(\hbar\bar{\omega}\epsilon_0)$ \cite{Duncan:2004,Lahaye:2009} 
(in trap units) with $E$ being an external electric field, $\epsilon_0$ the permittivity of free space, 
and $\alpha$ the static polarizability. Thus, the DDI can then be induced and tuned by an external field 
$E$ \cite{Yi:2000} to various orders of magnitude \cite{Bohn:2009} and in polar molecules the dipole 
moment can be set up to $10^4$ times larger than in atomic systems. The DDI also occur naturally 
\cite{Goral:2000,Olson:2013,Koch:2008,Lu:2011,Youn:2010} if the atoms possess a magnetic dipole moment 
$\bar{\mu}$ in which case $C_{dd}=\mu_0\bar{\mu}^2/(4\pi\hbar\bar{\omega})$ \cite{Lahaye:2009,Koch:2008,Kumar:2015} 
(in trap units) with $\mu_0$ the permittivity of free space. The control of DDI has also been further 
explained in the article by Lahaye \ea\ \cite{Lahaye:2009}. Moreover, by using the linear Stark effect, 
it is possible to excite Rydberg-dressed atoms to very high principle quantum numbers to achieve large 
dipole moments like $p\sim 1450D$, where $D$ is the ratio between the dipolar and $s$-wave scattering 
length \cite{Filinov:2012}. In this regard, we justify the use of large values of the DDI parameter 
\Gkdd\ in the present work. Before we continue, it should be emphasized that the DDI in the present 
work is repulsive.

\subsection{Mean-Field Gross-Pitaevskii Equation}

\hs The trapped DBEC is described by the time-dependent Gross-Pitaevskii equation (TDGPE). 
It is reduced from the 3D to the 1D form by integrating out the contributions in the transverse 
direction (see e.g. Refs.\ft\cite{Muruganandam:2009,Kumar:2015}). We thus consider a cigar-shaped DBEC
that is elongated along the $x$--axis with strong radial confinement in the transverse direction. 
The dynamics in the transverse direction is frozen in the radial ground state

\begin{equation}
\phi(\mathbf{\rho})\,=\,\frac{1}{d_\rho \sqrt{\pi}} e^{-\rho^2/(2 d_\rho^2)},
\label{eq:radial-ground-state}
\end{equation}

where $d_\rho$ is the width of the Gaussian. The reduced 1D TDGPE reads then as in Ref.\cite{Kumar:2015}

\begin{eqnarray}
&&i\frac{\partial\psi(x,t)}{\partial t}\,=\,\left[-\frac{1}{2}\frac{\partial^2}{\partial x^2}\,+\,
V_{tr}(x)\,+\,V_{DT}(x,t)\,+\,\right.\nonumber\\
&&\left. {\cal G}_{1D}|\psi(x,t)|^2\,+\,\right.\nonumber\\
&&\left. \frac{4\pi}{3} {\cal G}_{1dd} \int_{-\infty}^{+\infty} \frac{dk_x}{2\pi} 
e^{-i k_x x}|\widetilde{\psi}(k_x,t)|^2 j_{1D}(\tau_x)\right] \psi(x,t),\nonumber\\
\label{eq:one-dimensional-TDGPE}
\end{eqnarray}

where $\widetilde{\psi}(k_x,t)$ is the Fourier transform of $\psi(x,t)$. The latter is 
the longitudinal wave function which is normalized to one vis-$\acute{a}$-vis 
$\int_{-\infty}^{+\infty}|\psi(x,t)|^2 dx=1$ with $\psi(x,t)$ in units of $a_{ho}^{-1/2}$.  
$j_{1D}(\tau_x)$ is a function given by

\begin{equation}
j_{1D}(\tau_x)\,=\,\frac{\sqrt{2}}{2\pi d_\rho}\int_{-\infty}^{+\infty} d\tau_y e^{-\tau_y^2} h_{2D}(\tau),
\label{eq:Eq42inKumar2015}
\end{equation}

with $\tau_y\,=\,d_\rho k_y/\sqrt{2}$, $\tau\,=\,\sqrt{\tau_x^2\,+\,\tau_y^2}$, and 

\begin{equation}
h_{2D}(\tau)\,=\,\frac{1}{\sqrt{2\pi} d_\rho} \left[2\,-\,3\sqrt{\pi} e^{\tau^2} \tau\left\{1\,-\,erf(\tau)\right\}\right].
\end{equation}

${\cal G}_{1D}$ and \Gkdd\ are respectively the 1D s-wave and dipolar interaction parameters 
(see Sec.\ft\ref{sec:parameters} below). The fourth term on the right-hand-side of 
(\ref{eq:one-dimensional-TDGPE}) introduces the usual mean-field s-wave interaction nonlinearity. 
The last term introduces the mean-field dipolar nonlinearity derived from the long-range DDI.
It is a special integral that is designed to eliminate the singularity in the DDI potential. 

\hs Equation (\ref{eq:one-dimensional-TDGPE}) is solved numerically using the famous split-step 
Crank-Nicolson (CN) method \cite{Muruganandam:2009,Kumar:2015} in real time. It is first solved 
in imaginary time to initialize the BEC in the trapping geometry (\ref{eq:one-dimensional-external-power-law-trap}) 
superimposed on which is the NGP (\ref{eq:NGP-with-osc-depth}) without the modulated part, that is

\begin{equation}
V_{DT}^{(0)}(x)\,=\,A \exp(-\beta x^2).
\label{eq:VDT}
\end{equation}

In the second step, the DBEC is driven by the NGP (\ref{eq:NGP-with-osc-depth}) in real time to 
examine its ensuing dynamics. For further technical details, the reader is referred to Appendix 
\ref{sec:numerics}. The codes used for solving the TDGPE have been written by the group
of Antun Balaz in Belgrade and are fully explained in Ref.\cite{Kumar:2015} for recent versions 
treating DBECs and also Ref. \cite{Muruganandam:2009} for earlier versions on ordinary BECs. Numerous 
other TDGPE codes are available by this group \cite{Vudragovic:2012,Loncar:2016,
Loncar:2016b,Young:2016,Sataric:2016,Young:2017,Kumar:2019} and have been used extensively.

\subsection{Gross-Pitaevskii Energies}

\hs The Gross-Pitaevskii (GP) energies are as usual evaluated via

\begin{eqnarray}
&& E_{GP}(t)=\int_{-\infty}^{+\infty} dx\left[\,\left|\frac{\partial\psi(x,t)}{\partial x}\right|^2\,+\,\right.\nonumber\\
&&\left.\left[V_{tr}(x)\,+\,V_{DT}(x,t)\right]|\psi(x,t)|^2\,+\,\frac{1}{2} {\cal G}_{1D}|\psi(x,t)|^4\,+\,\right.\nonumber\\
&&\left.\frac{2\pi}{3} {\cal G}_{1dd}\int_{-\infty}^{+\infty}\frac{dk_x}{2\pi} e^{-ik_x x}|
\widetilde{\psi}(x,t)|^2 j_{1D}(\tau_x)|\psi(x,t)|^2\right].\nonumber\\
\end{eqnarray}

The time average is then computed using 

\begin{equation}
\langle E_{GP}\rangle_t=\frac{1}{T}\int_0^T E_{GP}(t)dt.
\label{eq:EGP}
\end{equation}

\subsection{Parameters}\label{sec:parameters}

\hs The 3D $s$-wave and dipolar interaction parameters are defined as ${\cal G}=4\pi N a$ and 
${\cal G}_{dd}=3 N a_{dd}$, where $a$ and $a_{dd}$ are the s-wave and dipolar scattering lengths, 
and $N$ the number of particles. The interaction parameters acting in 1D here, ${\cal G}_{1D}$ 
and ${\cal G}_{1dd}$, are obtained after the reduction of the 3D TDGPE to the 1D form 
(\ref{eq:one-dimensional-TDGPE}). As a result of the latter, the ${\cal G}$ and ${\cal G}_{dd}$ 
are divided by a factor $2\pi d_\rho^2$ so that

\begin{equation}
{\cal G}_{1D}\,=\,\frac{{\cal G}}{2\pi d_\rho^2} \hspace{0.5 cm} {\rm and} \hspace{0.5 cm} 
{\cal G}_{1dd}\,=\,\frac{{\cal G}_{dd}}{2 \pi d_\rho^2},
\label{eq:definition-of-G1D}
\end{equation}

where $d_\rho$ is the width of the integrated-out wave function in the transverse direction. 
The ${\cal G}_{1D}$ and \Gkdd\ are input directly into the code without explicit evaluation 
via $N$, $a$, $a_{dd}$, and $d_{\rho}$. These parameters define the strength of the s-wave
and dipolar-interaction nonlinearities of the TDGPE. The systems are simulated from $x=-L_0$ 
to $x=+L_0$ with $L_0=30$ for a harmonic oscillator (HO) trap and quartic trap (QT), and $L_0=51.2$ 
for a box potential. The parameters of the oscillating NGP are $A=-30$, $\beta=4$, and $\Omega=2\pi f$ 
with $f=10$. The values of $\delta A$ applied are 5, 10, and 20. The DBECs are scanned along a 
range of \Gkdd\ ranging from $0$ to $400$ in steps of 2 and for each \Gkdd\ a run was performed. 
Values of ${\cal G}_{1D}\,=\,50, 100, 150$ were used for each set of the runs. ${\cal G}_{1D}$, 
\Gkdd\, $L_0$, $a$, $a_{dd}$, and $d_\rho$ are all in units of \aho.

\subsection{Signal Energy}

\hs For a time-dependent physical observable $f(t)$, the signal energy ${\cal D}$ is 
formally defined by the integral 

\begin{equation}
{\cal D} =\int_{-\infty}^{+\infty} |f(t)|^2 dt,
\label{eq:Energy-spectral-density-definition}
\end{equation}

where $f(t)$ is taken to be the mean-field kinetic energy

\begin{equation}
E_{kin}(t)=\frac{1}{2}\int_{-\infty}^{+\infty}\bigg|\frac{\partial\psi{(x,t)}}{\partial x}\bigg|^2dx,
\label{eq:MFEkin}
\end{equation}

$\psi(x,t)$ being the time-dependent wavefunction describing the DBEC. We refer to the signal 
energy by ${\cal D}_{kin}$ because it is derived from $E_{kin}$. $E_{kin}$ is in units of 
\hw\ and ${\cal D}_{kin}$ in $(\hbar\bar{\omega})^2\bar{\omega}$. However, since we cannot 
numerically integrate to infinity, neither timely nor spatially, we limit the integral in
Eq.(\ref{eq:Energy-spectral-density-definition}) from $t=0$ to some time $T$ that is sufficiently long 
to reveal enough of the dynamical properties, and in Eq.(\ref{eq:MFEkin}) to the length of the simulation grid from 
$x=-L_0$ to $L_0$. The usefulness of the signal energy in displaying important properties about excitations in a 
driven BEC has already been demonstrated in our recent work \cite{Sakhel:2020}. There, it has 
been argued that the signal energy is tantamount to the time-average of the squared amplitude of 
an oscillating signal describing a dynamic variable. It can be likened to the root-mean-squared 
value of an alternating voltage or current. Since $E_{kin}(t)$ is found here again to be oscillating 
with time, it is quite convenient to apply the latter concept to its measurement. The reason for 
using the kinetic energy stems additonally from the fact that it is an important property of the condensate 
\cite{Nikitin:2005}. Other quantities, such as the potential energy, zero-point energy, and the 
radial size could also have been used here because they reveal the same PRs with the same properties 
as obtained for $E_{kin}$. Further, $E_{kin}$ has been used in a number of previous examinations 
\cite{Sakhel:2013,Sakhel:2016,Sakhel:2016b,Sakhel:2017,Sakhel:2018,Sakhel:2019} that further 
demonstrate its importance.

\hs As far as the experimental measurement of ${\cal D}_{kin}$ is concerned, this can be
performed as follows. While the BEC is excited by the focusing red-laser beam of an oscillating
intensity, an in-situ recording of density profiles $n(x,t)$ is performed as a function of time 
by a CCD camera using a nondestructive method as that demonstrated in the experiment of Onofrio 
\ea\ \cite{Onofrio:2000}. The latter performed a repeated in-situ non-destructive phase-contrast 
imaging that has been used by Andrews \ea\ \cite{Andrews:1996} and Ketterle \ea\ \cite{Ketterle:1999}. 
After the $n(x,t)$ are recorded, they are Fourier-transformed to obtain the corresponding 
momentum-density distributions $n(\mathbf{p},t)$ that can be used to compute $E_{kin}(t)$ via

\begin{equation}
E_{kin}(t)\,=\,\frac{1}{2m}\int dp\,p^2 n(\mathbf{p},t).
\label{eq:Ekin-via-momentum-density}
\end{equation}

Having obtained $E_{kin}$, ${\cal D}_{kin}$ is then evaluated easily.

\begin{figure}
\includegraphics[width=8.5cm,bb=130 214 476 739,clip]{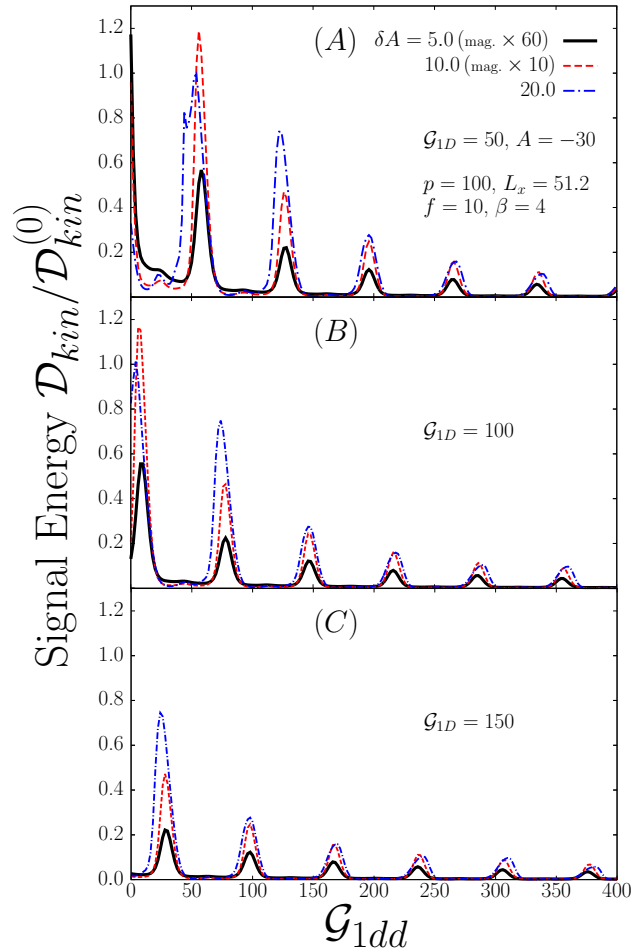}
\caption{Signal energy ${\cal D}_{kin}$ of a driven one-dimensional DBEC in a box as a function of 
the dipolar interaction parameter \Gkdd\ [Eq.(\ref{eq:definition-of-G1D})]. ${\cal D}_{kin}$ is normalized 
by ${\cal D}^{(0)}_{kin}$, the maximum of ${\cal D}_{kin}$ at $\delta A=20$ in the ${\cal G}_{1dd}$-range 
considered. ${\cal G}_{1D}$ is the $s-$wave interaction parameter [Eq.(\ref{eq:definition-of-G1D})]. 
The box is generated from Eq.(\ref{eq:one-dimensional-external-power-law-trap}) using $L_x=51.2$ and $p=100$. 
The DBEC is excited by an NGP whose principal depth $A$ is periodically modulated by an amplitude $\delta A$ 
via Eq.(\ref{eq:NGP-with-osc-depth}) with $A=-30$, $f=10$, and $\beta=4$. Various values 
are used for $\delta A$ and \GkD. Frame ($A$) and thick-solid line: ${\cal D}_{kin}$ at ${\cal G}_{1D}=50$ 
with $\delta A=5$. The signal is magnified (mag.) by 60 times ($\times$) to enable comparison with 
other cases; thin-dashed line: $\delta A=10.0$ (mag.$\times 10$); and dashed-dotted line: 20.0 (no mag.). 
Frames ($B,C$): as in ($A$); but for ${\cal G}_{1D}=100$ and 150, respectively. $A$ and $\delta A$ are 
in units of \hw; $L_x$, ${\cal G}_{1D}$, \Gkdd\ in \aho, whereas $\beta$ 
in $a_{ho}^{-2}$. ${\cal D}_{kin}$ and ${\cal D}^{(0)}_{kin}$ are in $(\hbar\bar{\omega})^2\bar{\omega}$.}
\label{fig:plotDataEFTSignalvsGdipG1D50A30B4Powx100Lx100Wext0Freq10sevdA}
\end{figure}

\begin{figure}
\includegraphics[width=8.5cm,bb=160 314 450 739,clip]{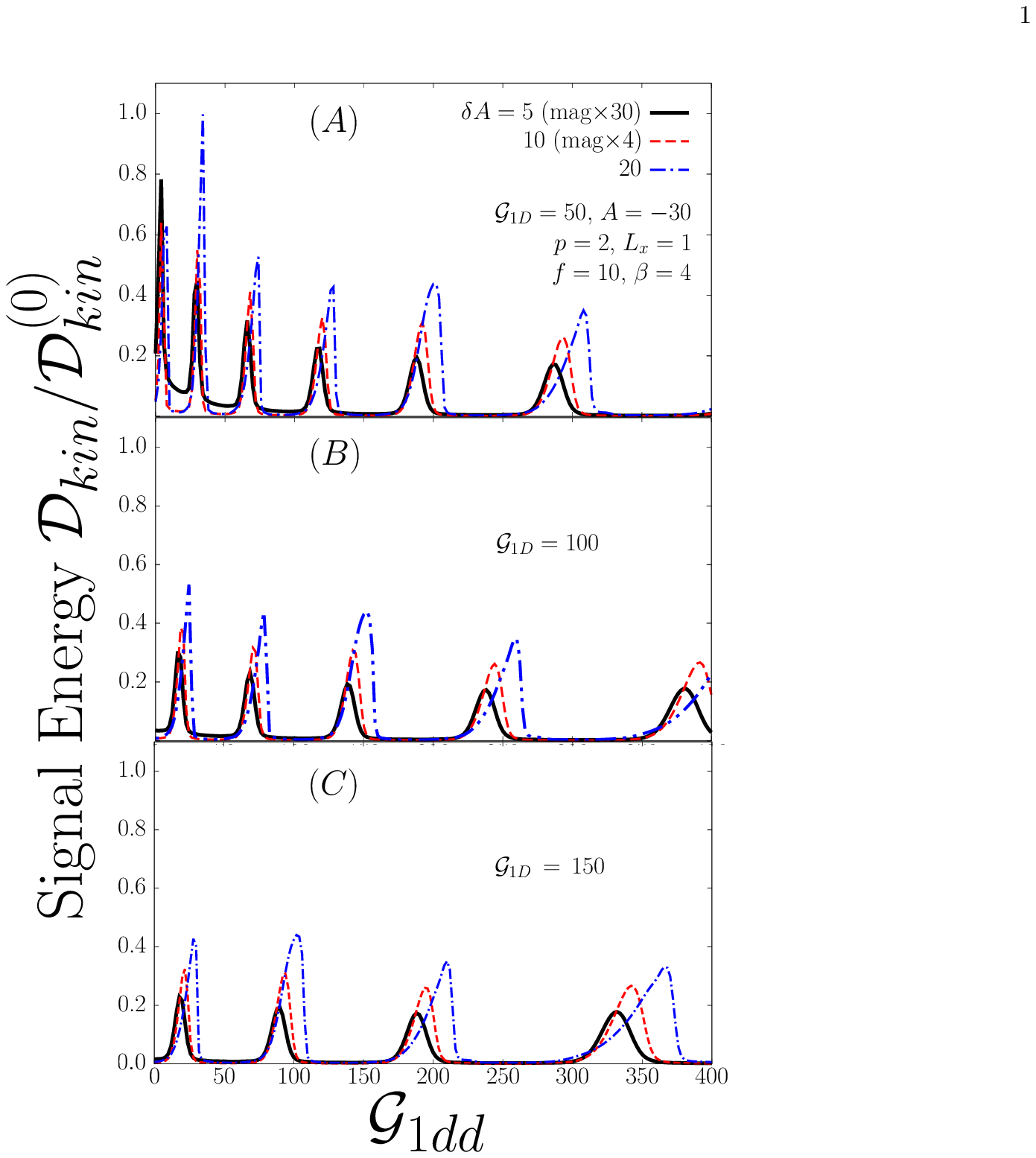} 
\caption{As in Fig.\ft\ref{fig:plotDataEFTSignalvsGdipG1D50A30B4Powx100Lx100Wext0Freq10sevdA}; but 
for an HO trap with $p=2$, $L_x=1$, magnifications of $\times 30$ for $\delta A=5$ and $\times 4$ for 
$\delta A=10$, and ${\cal D}_{kin}^{(0)}$ is the maximum of \Dkin\ at $\delta A=20$ in frame ($A$). 
$A$ and $\delta A$ are in units of \hw; $L_x$, ${\cal G}_{1D}$, \Gkdd\ in \aho, whereas $\beta$ in 
$a_{ho}^{-2}$. ${\cal D}_{kin}$ and ${\cal D}^{(0)}_{kin}$ are in $(\hbar\bar{\omega})^2\bar{\omega}$.}
\label{fig:plotDataEFTSignalvsGdipG1DA30B4Powx2Lx1Wext0Freq10sevdA}
\end{figure}

\section{Results}\label{sec:results}

\subsection{Resonances in a Box Potential}\label{sec:resonances-in-a-box}

\hs Figure\ft\ref{fig:plotDataEFTSignalvsGdipG1D50A30B4Powx100Lx100Wext0Freq10sevdA} displays 
${\cal D}_{kin}/{\cal D}^{(0)}_{kin}$ vs \Gkdd\ for a few different values of \GkD\ and $\delta A$ 
in a BOX$+$NGP trap. ${\cal D}^{(0)}_{kin}$ is the maximum signal energy for ${\cal G}_{1D}=50$ at 
$\delta A=20$ in the range of \Gkdd\ considered and is used for normalization of the signals in all 
frames. Several PRs are discovered whose amplitudes decline with increasing values of \Gkdd. This is 
in line with Refs.\cite{Olson:2013,Mendonca:2018}, where it has been found that the DDI 
reduce the amplitude of DBEC dynamics. An increase in the DDI causes therefore a weaker response 
to the oscillating NGP as it turns out that the repulsive DDI reduce the depth of the effective 
mean-field potential (see Sec.\ft\ref{sec:effective-potential} below) and with it the occupancy of 
the NGP so that lesser particles are excited.

\hs Qualitatively, the same features are observed for all values of \GkD\ in 
Fig.\ft\ref{fig:plotDataEFTSignalvsGdipG1D50A30B4Powx100Lx100Wext0Freq10sevdA}. Notably, an increase 
in \GkD\ shifts the whole spectrum backwards (to the left) keeping the same distance $\Delta {\cal G}_{1dd}$ 
between each pair of peaks, demonstrating that \dGkdd\ is not influenced by \GkD. For example, the third 
peak from the left in frame ($A$) at \Gkdd\ $\sim 200$ is shifted backwards by an amount of 50 in frame ($B$) from its 
position in ($A$), and by an amount 100 in frame ($C$). That is, an increase of \GkD\ by 50 causes all
the resonance positions to shift backwards by the same value of 50, (similarly for an increase by 100) 
being a rather unprecedented and remarkable feature. Thus there seems to be no effect for the interplay 
between $s$-wave interactions and DDI on the principle features of the spectrum of ${\cal D}_{kin}$ in 
a box potential. This demonstrates that $s$-wave interactions and DDI work similarly in determining the 
positions of PRs. The above results reveal information about the energy-level structure of the DBEC and 
shape of the trapping potential. The equidistance of peaks mirrors the confinement homogeneity, i.e. the 
flatness of the box. This can be particularly concluded from a comparison with the HO potential in the 
next section.

\hs The reason for the PRs and the shift of their positions with ${\cal G}_{1D}$ are discussed in 
Sec.\ft\ref{sec:origins-of-PRs} basing on well-known theoretical methods: the Wenzel-Kramer-Brillouin (WKB) 
approximation, time-independent and time-dependent perturbation theory, and LVM. Most importantly, it is shown that the 
PRs arise whenever the time-averaged energy of the DBEC closely matches one of the energy levels of 
the trap$+$NGP.

\subsection{Resonances in a Harmonic Trap}\label{sec:resonances-in-a-HO-trap}

\hs In Fig.\ft\ref{fig:plotDataEFTSignalvsGdipG1DA30B4Powx2Lx1Wext0Freq10sevdA}, PRs 
are also encountered in an HO$+$NGP trap, except that the separation \dGkdd\ is not uniform as 
in the box but rather increases with \Gkdd. Henceforth, this mirrors confinement inhomogeneity. 
Further, this reveals the role of the interplay between 
trapping geometry and DDI in determining the PR energies. A uniform trap leads to equidistant 
whereas a nonuniform one to non-equidistant PRs along the \Gkdd\ axis. What is peculiar though, is that the 
shift of the whole spectrum as a result of changing ${\cal G}_{1D}$ is also observed here in the same uniform 
manner as reported in Fig.\ft\ref{fig:plotDataEFTSignalvsGdipG1D50A30B4Powx100Lx100Wext0Freq10sevdA}. 
Increasing ${\cal G}_{1D}$ by 50 causes the PR peaks to shift backwards by 50 along 
the \Gkdd\ axis.

\subsection{Resonances in a Quartic Trap}\label{sec:resonances-in-a-Quartic-Trap}

\begin{figure}
\includegraphics[width=8.7cm,bb=160 306 456 739,clip]{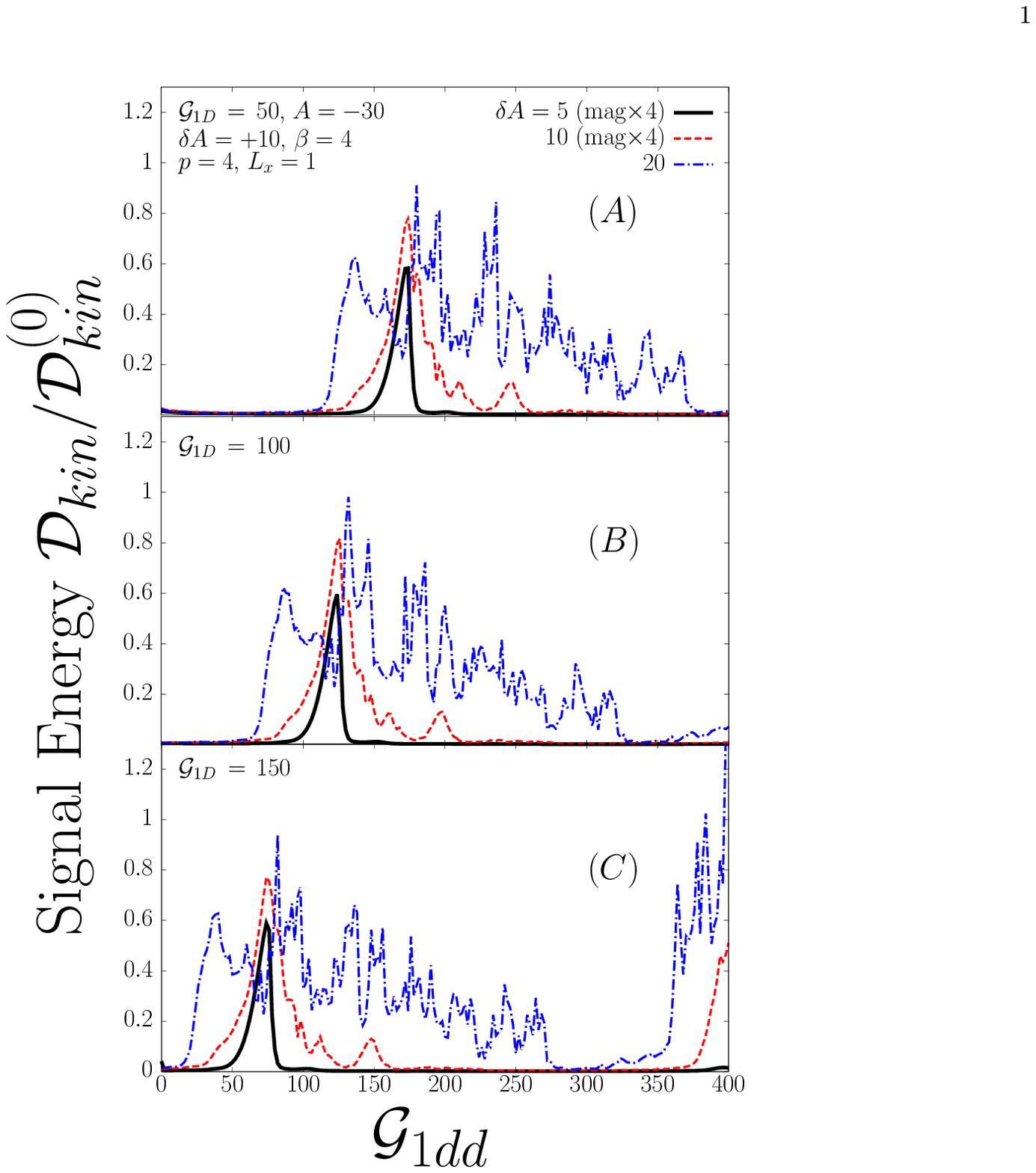} 
\caption{As in Fig.\ft\ref{fig:plotDataEFTSignalvsGdipG1D50A30B4Powx100Lx100Wext0Freq10sevdA}; 
but for a QT with $p=4$, $L_x=1$, magnification of $\times 4$ for $\delta A=5$ and 10, and 
${\cal D}_{kin}^{(0)}$ is a normalization factor. $A$ and $\delta A$ are in units of \hw; $L_x$, ${\cal G}_{1D}$, 
\Gkdd\ in \aho, whereas $\beta$ in $a_{ho}^{-2}$. ${\cal D}_{kin}$ and ${\cal D}^{(0)}_{kin}$ are in 
$(\hbar\bar{\omega})^2\bar{\omega}$}
\label{fig:plotDataEFTSignalvsGdipG1DA30B4Powx4Lx1Wext0Freq10sevdA}
\end{figure}

\hs In Fig.\ft\ref{fig:plotDataEFTSignalvsGdipG1DA30B4Powx4Lx1Wext0Freq10sevdA} for a QT$+$NGP, 
there is only one well-defined PR at $\delta A=5$ in the same range of \Gkdd\ considered. 
At the larger $\delta A$, a disordered excitation pattern arises. Compared to 
Figs.\ft\ref{fig:plotDataEFTSignalvsGdipG1D50A30B4Powx100Lx100Wext0Freq10sevdA} 
and \ref{fig:plotDataEFTSignalvsGdipG1DA30B4Powx2Lx1Wext0Freq10sevdA}, this is rather 
surprising since a pattern similar to that for the HO$+$NGP trap had been anticipated. Thus 
for a QT at stronger driving quite a larger number of modes is excited than in the HO 
trap and box. As such, there exist trapping geometries that under certain conditions 
do not support ordered PR patterns. Once again it can be seen that the geometry of the trap 
strongly influences the PR phenomenon and its pattern. Therefore, it can be used to control the 
PRs. As in the previous figures, an increase of ${\cal G}_{1D}$ by 50 causes the whole PR spectrum 
to shift by 50 along the ${\cal G}_{1dd}$ axis.

\begin{figure}
\includegraphics[width=8.5cm,bb=186 440 429 742,clip]{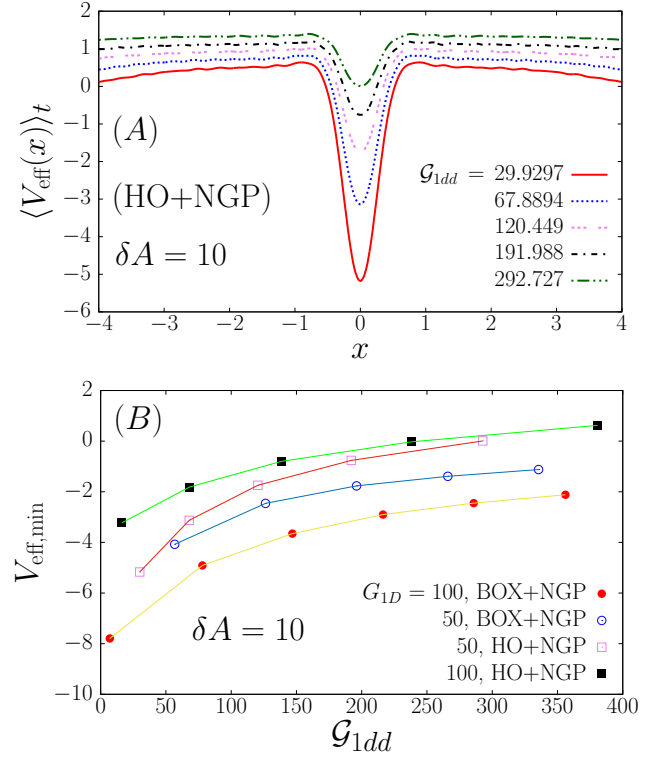}
\caption{Frame ($A$): Time-averaged effective potential $\langle V_{eff}(x)\rangle_t$ 
[Eq.(\ref{eq:effectivepotential-time-averaged})] for the HO+NGP in 
Fig.\ft\ref{fig:plotDataEFTSignalvsGdipG1DA30B4Powx2Lx1Wext0Freq10sevdA}($A$) at $\delta A=10$ and the 
indicated resonant values of \Gkdd. Solid line: 29.9297; dotted line: 67.8894; triple-dotted line: 120.449;
dashed-dotted line: 191.988; dashed-double-dotted line: 292.727. Frame ($B$): Minima 
$V_{\tiny\hbox{eff,min}}=\langle V_{eff}(0)\rangle_t$  at the PR values of ${\cal G}_{1dd}$ for some 
of the systems presented in Figs.\ft\ref{fig:plotDataEFTSignalvsGdipG1D50A30B4Powx100Lx100Wext0Freq10sevdA}  
and \ref{fig:plotDataEFTSignalvsGdipG1DA30B4Powx2Lx1Wext0Freq10sevdA} at $\delta A=10$: Solid circles: 
BOX+NGP at ${\cal G}_{1D}=100$; open circles: same at 50; open squares: HO+NGP at ${\cal G}_{1D}=50$; 
solid squares: same at 100. $\langle V_{eff}(x)\rangle_t$, $V_{\tiny\hbox{eff,min}}$, $A$, and 
$\delta A$ are in units of \hw, whereas $L_x$, $x$, ${\cal G}_{1D}$, and \Gkdd\ in units of \aho, and 
$\beta$ in $a_{ho}^{-2}$.}
\label{fig:plotEffectivePotentialsInformationFigureStack}
\end{figure}

\subsection{Effective Potential}\label{sec:effective-potential}

\hs To this end, it is useful to state the reason for the decline of the amplitude of a PR with 
${\cal G}_{1dd}$ as observed in Figs.\ft\ref{fig:plotDataEFTSignalvsGdipG1D50A30B4Powx100Lx100Wext0Freq10sevdA} 
and \ref{fig:plotDataEFTSignalvsGdipG1DA30B4Powx2Lx1Wext0Freq10sevdA}. This decline is attributed to a change in the 
depth of the time-averaged effective potential given by  

\begin{eqnarray}
&&\langle V_{eff}(x)\rangle_t=\Bigg\langle V_{DT}(x,t)\,+\,{\cal G}_{1D}|\psi(x,t)|^2\,+\,\nonumber\\
&&\frac{4\pi}{3}{\cal G}_{1dd}\int_{-\infty}^{+\infty} \frac{dk_x}{2\pi} e^{-i k_x x} j_{1D}(\tau_x)
|\widetilde{\psi}(k_x,t)|^2 \Bigg\rangle\,+\,V_{tr}(x),\nonumber\\
\label{eq:effectivepotential-time-averaged}
\end{eqnarray}

with 

\begin{equation}
\langle\cdots\rangle=\frac{1}{T}\int_0^T(\cdots)dt,
\end{equation}

$T$ being the total simulation time. Fig.\ft\ref{fig:plotEffectivePotentialsInformationFigureStack}($A$) shows as an example 
$\langle V_{eff}(x)\rangle_t$ for the HO+NGP of Fig.\ft\ref{fig:plotDataEFTSignalvsGdipG1DA30B4Powx2Lx1Wext0Freq10sevdA}($A$). 
Fig.\ft\ref{fig:plotEffectivePotentialsInformationFigureStack}($B$) 
displays the minimum of $\langle V_{eff}(x)\rangle_t$ at $x=0$ for previous systems in 
Figs.\ft\ref{fig:plotDataEFTSignalvsGdipG1D50A30B4Powx100Lx100Wext0Freq10sevdA} 
and \ref{fig:plotDataEFTSignalvsGdipG1DA30B4Powx2Lx1Wext0Freq10sevdA}
as a function of the PR values of \Gkdd. $\langle V_{eff}(x)\rangle_t$ 
becomes shallower as \Gkdd\ rises causing lesser particles to occupy the NGP. Hence, lesser 
particles contribute to the strength of the excitations as they are thrown out of the NGP by 
its oscillating depth thereby causing the drop in the PR amplitudes ${\cal D}_{kin}$. Notably, 
$\langle V_{eff}(x)\rangle_t$ closely follows the spatial shape of the NGP and it can be therefore 
argued that the influence of the NGP on the DBEC is reduced as $\langle V_{eff}(x)\rangle_t$ 
becomes shallower. Further the depth $\langle V_{eff}(x)\rangle_t$ being controlled by \Gkdd\ 
plays also a decisive role in determining the PR energies. 


\section{Origins of the parametric resonances}\label{sec:origins-of-PRs}

\hs The reasons for the appearance of the PRs at certain values of 
\Gkdd\ are deeply encrypted in the numerical solutions of the TDGPE and hard to decipher. 
Therefore, there is no other way to circumvent this problem but to seek qualitative explanations 
basing on other models and methods, such as LVM \cite{Muruganandam:2012,Sakhel:2017,
Sakhel:2018,Perez-Garcia:1996,Perez-Garcia:1997,Wang:2002,Al-Jibbouri:2013,
Falco:2009,Wei:2013}, the WKB approximation, as well as time-dependent and time-independent 
perturbation theory \cite{Merzbacher:1998,Griffiths:2018}. It should be noted that LVM has been 
applied earlier to treat the same HO+NGP system as the present one, but without DDI \cite{Sakhel:2018}. 
The latter investigation also demonstrated the presence of PRs although in a different framework. 
Astonishingly, by using LVM in the present work it has been possible to generate a few resonances in a 
manner similar to those obtained in Figs.\ft\ref{fig:plotDataEFTSignalvsGdipG1D50A30B4Powx100Lx100Wext0Freq10sevdA}--
\ref{fig:plotDataEFTSignalvsGdipG1DA30B4Powx4Lx1Wext0Freq10sevdA}. The LVM analysis of these PRs can be used 
in a qualitative manner to cast more light on them and to reach a better understanding of their origin. 
The time-averaged GP energies of some of our systems at PR were matched to energy levels calculated 
by using the perturbation theory corresponding to either the HO$+$NGP, BOX$+$NGP, or QT$+$NGP traps. 
This enables us to characterize their level structure. By applying time-dependent perturbation theory, 
it has been demonstrated that each PR corresponds to a certain transition between an energy level $m$ 
of the trap and the energy level $n$ to which the PR has been matched, identified by the maximum transitional 
probability between a set of other probabilities to this state $n$.

\subsection{LVM analysis}

\subsubsection{Euler-Lagrange Differential Equation}

\hs The Euler-Lagrange equation for the dynamics of the width $w(t)$ of the DBEC in 
a 1D harmonic trap can be found in Ref.\cite{Muruganandam:2012}. We thus add to this LVM 
equation the contribution coming from the NGP with oscillating depth [Eq.(\ref{eq:NGP-with-osc-depth})]. 
Further, using our definitions for \Gkdd\ and \GkD\ [Eq.(\ref{eq:definition-of-G1D})], 
this LVM equation becomes

\begin{eqnarray}
&&\ddot{w}\,+\,w\,=\,\frac{1}{w^3}\,+\,\frac{{\cal G}_{1D}}{\sqrt{2\pi} w^2}\,-\,\nonumber\\
&&\frac{\beta [A+\delta A \cos(\Omega t)] w}{(1\,+\,\beta w^2)^{3/2}}\,-\,2\sqrt{2\pi}\frac{{\cal G}_{1dd}}{3 w^2} c(\kappa)
\label{eq:Euler-Lagrange-Equation-LVM-Muruganandam}
\end{eqnarray}

with $\kappa\,=\,d_\rho/w$,

\begin{equation}
c(\kappa)\,=\,\frac{1\,+\,10 \kappa^2\,-\,2 \kappa^4\,-\,9 \kappa^2 d(\kappa)}{(1\,-\,\kappa^2)^2},
\label{eq:c(k)-dipolar-interaction}
\end{equation}

and

\begin{equation}
d(\kappa)\,=\,\frac{\hbox{arctanh}\sqrt{1-\kappa^2}}{\sqrt{1-\kappa^2}}.
\label{eq:d(k)-dipolar-interaction}
\end{equation}

The width at time $t=0$ is used as an initial condition for solving the LVM 
equation (\ref{eq:Euler-Lagrange-Equation-LVM-Muruganandam}) and is taken to be the value 
$w_0=w(t=0)$ at which the DBEC is in equilibrium. As in Refs.\cite{Sakhel:2017,Sakhel:2018}, 
$w_0$ is obtained by solving the equation

\begin{equation}
w_0\,-\,\frac{1}{w_0^3}\,-\,\frac{{\cal G}_{1D}}{\sqrt{2\pi} w_0^2}\,+\,
\frac{A \beta w_0}{(1\,+\,\beta w_0^2)^{3/2}}\,+
\,2\sqrt{2\pi}\frac{{\cal G}_{1dd}}{3 w_0^2} c(\kappa_0)\,=\,0,
\label{eq:Euler-Lagrange-Equation-LVM-Muruganandam-at-equilibrium}
\end{equation}

at $\kappa_0\,=\,d_\rho/w_0$ that is gotten from Eq.(\ref{eq:Euler-Lagrange-Equation-LVM-Muruganandam}
by setting $\delta A=0$ and $\ddot{w}=0$.

\subsubsection{Frequency of Breathing Mode}

\hs In this section, we revisit the frequency of the breathing mode 
explored earlier in some of our publications \cite{Sakhel:2017,Sakhel:2018},
except that this time the effects of the DDI are added. As before, the breathing-mode 
frequency can be obtained by a linearization of Eq.(\ref{eq:Euler-Lagrange-Equation-LVM-Muruganandam}) 
via $w(t)=w_0+\delta w(t)$. However, to avoid mathematical complexities, it
is easier to equivalently take the differential of both sides of 
(\ref{eq:Euler-Lagrange-Equation-LVM-Muruganandam}) with respect to $w$ at
$w_0$. Consequently, one gets

\begin{equation}
\delta \ddot{w}\,+\,\Omega_B^2\delta w\,=\,0
\end{equation}

with $\Omega_B^2$ given by

\begin{eqnarray}
\Omega_B^2(t)\,&=&\,1\,+\,\frac{3}{w_0^4}\,+\,\sqrt{\frac{2}{\pi}}\frac{{\cal G}_{1D}}{w_0^3}\,+\,\nonumber\\
&&\frac{[A + \delta A \cos(\Omega t)] \beta (1\,-\,2 \beta w_0^2)}{(1\,+\,\beta w_0^2)^{5/2}}\,+\,\nonumber\\
&&2\sqrt{2\pi}\frac{{\cal G}_{1dd}}{3 w^2}\left[\frac{\partial c(\kappa)}{\partial w}\Bigg|_{w_0}\,-\,
\frac{2}{w} c(\kappa_0)\right].
\label{eq:omegaB-breathing-frequency}
\end{eqnarray}

The term between brackets yields

\begin{eqnarray}
&&\frac{\partial c(\kappa)}{\partial w}\Bigg|_{\kappa_0}\,-\,\frac{2}{w_0} c(\kappa_0)\,=\,\nonumber\\
&&\frac{\kappa_0}{d_\rho (1\,-\,\kappa_0^2)^3}\Bigg(-2\,-\,51\kappa_0^2\,+\,12\kappa_0^4\,-\,4\kappa_0^6\,+\,\nonumber\\ 
&& \frac{9\kappa_0^2(\kappa_0^2\,+\,4)}{\sqrt{1\,-\,\kappa_0^2}} \hbox{arctanh}\sqrt{1\,-\,\kappa_0^2}\Bigg).
\end{eqnarray}

Again, it can be concluded that an imaginary value of $\Omega_B$ is obtained if

\begin{eqnarray}
&&1\,+\,\frac{3}{w_0^4}\,+\,\sqrt{\frac{2}{\pi}}\frac{{\cal G}_{1D}}{w_0^3}\,<\,\nonumber\\
&&-\frac{[A + \delta A\cos(\Omega t)] \beta (1\,-\,2 \beta w_0^2)}{(1\,+\,\beta w_0^2)^{5/2}}\,\nonumber\\
&&-\,2\sqrt{2\pi}\frac{{\cal G}_{1dd}}{3 w_0^2}\left[\frac{\partial c(\kappa)}{\partial w}\Bigg|_{w_0}\,-\,\frac{2}{w_0} c(\kappa_0)\right],
\end{eqnarray}

that leads to a damping of the DBEC oscillations.

\subsubsection{LVM kinetic energy}

\hs The LVM kinetic energy is given by \cite{Sakhel:2017,Sakhel:2018,Muruganandam:2012}

\begin{equation}
e_{kin}(t)\,=\,\frac{1}{2}[\dot{w}(t)]^2\,+\,\frac{1}{2[w(t)]^2},
\label{eq:LVM-kinetic-energy}
\end{equation}

and is used now for $f(t)$ in Eq.(\ref{eq:Energy-spectral-density-definition}) for evaluating 
${\cal D}_{kin}$ and checking the possibility of generating PRs by the LVM equation 
(\ref{eq:Euler-Lagrange-Equation-LVM-Muruganandam}).

\subsubsection{LVM Results}

\hs The LVM equation (\ref{eq:Euler-Lagrange-Equation-LVM-Muruganandam}) is solved numerically with 
Mathematica over the same range of ${\cal G}_{1dd}$ as in Sec.\ft\ref{sec:results} using similar 
parameters for ${\cal G}_{1D}$, $A$, $\delta A$, $L_x$, and $\beta$, except for $\Omega$ which is set to 
other values for the best response of the system to changes in ${\cal G}_{1dd}$. For example, 
an integer value for $\Omega$ yields resonant behavior in $w(t)$ and ${\cal D}_{kin}$ contrary to 
a real value \cite{Sakhel:2018}. In this regard, it should be emphasized that we principally aim at 
a qualitative comparison with TDGPE that would help us in explaining the PRs of Sec.\ft\ref{sec:results}.

\begin{figure}
\includegraphics[width=8.5cm,bb=67 456 539 742,clip]{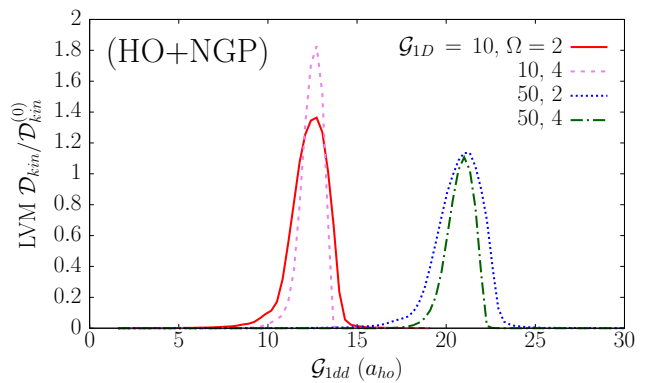}
\caption{Parametric resonances generated by the Euler-Lagrange differential equation 
(\ref{eq:Euler-Lagrange-Equation-LVM-Muruganandam}) for a 1D DBEC in the 
same HO$+$NGP trap of Fig.\ft\ref{fig:plotDataEFTSignalvsGdipG1DA30B4Powx2Lx1Wext0Freq10sevdA}. 
The graph displays the signal energy \Dkin\ of the LVM kinetic energy 
$e_{kin}(t)$ [Eq.(\ref{eq:LVM-kinetic-energy})] versus the DDI interaction parameter 
${\cal G}_{1dd}$. \Dkin\ is normalized by ${\cal D}_{kin}^{(0)}=1\times 10^6$. 
Different values for ${\cal G}_{1D}$ and $\Omega$ are considered: 10 and 2, 
respectively (solid line); 10, 4 (dashed line); 50, 2 (dotted line); and 50, 4 
(dashed-dotted line). The initial conditions used in solving the 
Euler-Lagrange equation (\ref{eq:Euler-Lagrange-Equation-LVM-Muruganandam}) at time 
$t=0$ are the initial ``speed" $\dot{w}\,=\,\dot{w}(0)\,=\,0$ and the values of the 
equilibrium $w_0\,=\,w(0)$ satisfying Eq.(\ref{eq:Euler-Lagrange-Equation-LVM-Muruganandam-at-equilibrium}) 
at each \Gkdd\ with the value of $d_\rho$ in $\kappa=d_\rho/w$ set to 0.6. ${\cal G}_{1D}$, ${\cal G}_{1dd}$, 
and $L_x$ are in units of $a_{ho}$ whereas $A$ in $\hbar\bar{\omega}$ and $\beta$ in $a_{ho}^{-2}$. 
\Dkin\ and \Dkino\ are in units of $(\hbar\bar{\omega})^2\bar{\omega}$.}
\label{fig:plotLVMekinvsG1ddsevG1Dpowx2}
\end{figure}

\hs In Fig.\ft\ref{fig:plotLVMekinvsG1ddsevG1Dpowx2}, the signal energy ${\cal D}_{kin}$ obtained 
from $e_{kin}(t)$, where the latter is given by (\ref{eq:LVM-kinetic-energy}), is graphed as a 
function of ${\cal G}_{1dd}$ and demonstrates that LVM surprisingly generates PRs along the same 
lines as the ones displayed in Sec.\ft\ref{sec:results}. Another astonishing result is that these 
resonances are shifted in their positions along the ${\cal G}_{1dd}$ axis when ${\cal G}_{1D}$ is 
changed. This is similar to what has been reported earlier in 
Figs.\ft\ref{fig:plotDataEFTSignalvsGdipG1D50A30B4Powx100Lx100Wext0Freq10sevdA}--
\ref{fig:plotDataEFTSignalvsGdipG1DA30B4Powx4Lx1Wext0Freq10sevdA} of the TDGPE results, except that 
in LVM they shift in the opposite direction and their amplitudes and shapes in 
Fig.\ft\ref{fig:plotLVMekinvsG1ddsevG1Dpowx2} change somewhat because of this shift. This opposite 
behavior is an artefact of the model arising from the fact that LVM relies on the variational 
Gaussian Ansatz \cite{Sakhel:2017,Muruganandam:2012,Perez-Garcia:1996} 

\begin{equation}
\psi(x)=\frac{1}{\pi^{1/4}w}e^{-\frac{x^2}{2w^2}+i\beta x^2}
\end{equation}

to evaluate the mean-field Lagrangian 
\cite{Muruganandam:2012,Sakhel:2017,Sakhel:2018,Perez-Garcia:1996,Perez-Garcia:1997,Al-Jibbouri:2013,Falco:2009,Wei:2013}. 
This Ansatz is much less flexible than the numerical solution to the TDGPE resulting in this different 
behavior. It should further be iterated that it is mostly suitable for BECs in an HO trap and thus needs to be 
developed for other types of traps. Nevertheless the discovery of the LVM PRs as presented in Fig.\ft\ref{fig:plotLVMekinvsG1ddsevG1Dpowx2} strongly substantiates 
our results on the TDGPE PRs and supports our claims that they are an inherent feature in the DBECs, and not just a result of some 
other influence such as noise or numerical chaos.

\begin{figure}
\includegraphics[width=8.5cm,bb=67 456 539 742,clip]{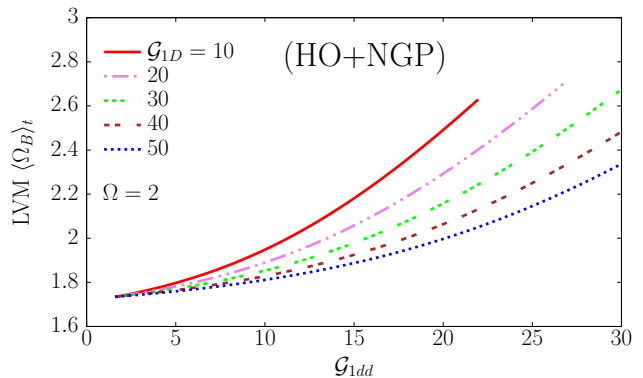}
\caption{As in Fig.\ft\ref{fig:plotLVMekinvsG1ddsevG1Dpowx2}; but for the time average 
$\langle\Omega_B\rangle_t$ of the LVM breathing-mode frequency $\Omega_B(t)$ given by  
[Eq.(\ref{eq:omegaB-breathing-frequency})] at various values of \GkD. Solid line: 
${\cal G}_{1D}=10$; dashed-double-dotted line: 20; triple-dotted line: 30; double-dotted line: 40;
dotted line: 50. $\Omega_B(t)$ is in units of $\bar{\omega}$.} \label{fig:plotLVMOmegavsG1ddsevG1Dpowx2}
\end{figure}

\hs Figure\ft\ref{fig:plotLVMOmegavsG1ddsevG1Dpowx2} displays the time-average of $\Omega_B(t)$ 
[Eq.(\ref{eq:omegaB-breathing-frequency})], that is $\langle\Omega_B\rangle_t$ graphed against 
${\cal G}_{1dd}$ for various ${\cal G}_{1D}$. For a larger ${\cal G}_{1D}$, the height of the curve drops. 
This can be connected to the PR shifting to larger resonant values of ${\cal G}_{1dd}$ 
in Fig.\ft\ref{fig:plotLVMekinvsG1ddsevG1Dpowx2}. Inspecting Eq.(\ref{eq:omegaB-breathing-frequency}), 
one can see that the term proportional to ${\cal G}_{1D}$ is linearly added to that proportional to 
${\cal G}_{1dd}$. Thus when $\langle\Omega_B \rangle_t$ is changed as a result of varying 
${\cal G}_{1D}$, the position of the corresponding PR is updated.

\begin{table*}
\caption{Match of the time averaged GP energies $\langle E_{GP}\rangle_t$ [Eq.(\ref{eq:EGP})] with the energy 
levels $E_n$ obtained from second-order perturbation theory. The system considered here is the 
HO+NGP trap in Fig.\ft\ref{fig:plotDataEFTSignalvsGdipG1DA30B4Powx2Lx1Wext0Freq10sevdA}($A$) 
for $\delta A=10$. The $E_n$ are obtained from Eqs.(\ref{eq:perturbative-first-order-correction}--\ref{eq:perturbative-total-energy-level}). 
From left to right the table lists \Gkdd, $\langle E_{GP}\rangle_t$, 
$E_n$ at quantum state $n$, the number of states $M$ found for the 
second-order correction (\ref{eq:perturbative-second-order-correction}), and the difference 
between the matched energies. $\langle E_{GP}\rangle_t$ and $E_n$ are in units of \hw, 
and ${\cal G}_{1dd}$ in \aho.}
\begin{tabular}{|*6{@{\hspace{0.5cm}} c @{\hspace{0.5cm}} |}} \hline
\Gkdd\ & $\langle E_{GP}\rangle_t$ &     $E_n$    & $n$ & $M$ & $|\langle E_{GP} \rangle_t\,-\,E_n|$ \\ \hline
      29.9297    &        2.67631            &   2.68257    &  4  & 10  &    0.00626 \\
      67.8894    &        5.79083            &   5.77971    &  7  & 20  &    0.01112 \\
      120.449    &        9.62129            &   9.5773     &  10 & 21  &    0.04399 \\
      191.988    &        15.1178            &   15.1395    &  13 & 11  &    0.02170 \\
      292.727    &        18.3912            &   18.3361    &  18 & 21  &    0.05510 \\ \hline\hline
\end{tabular}
\label{tab:HO+NGP-matched-to-GP-energies}
\end{table*}


\subsection{Energy Levels of the HO+NGP Trap via Perturbation Theory}\label{sec:perturbation-theory-energy-levels-HO}

\hs The goal of this section is to demonstrate that the GP time-averaged total energies 
$\langle E_{GP}\rangle_t$ [Eq.(\ref{eq:EGP})] at which the driven DBEC resonates can be generated 
from first and second-order perturbation theory thereby revealing a certain ELS  
via the quantum numbers $n$ corresponding to $\langle E_{GP}\rangle_t$. The HO$+$NGP is considered 
here first. Within this context, the NGP is taken to be a small perturbation within a large HO trap. 
This is in view of the fact that the size of our simulated system is 60 ($a_{ho}$) casting a very a 
high potential energy at the trap edges of the order of $\sim 450\,(\hbar\omega)$. Therefore, it is 
reasonable to apply time-independent perturbation theory to calculate the energy levels of the 
DBEC. The first order correction to the energy is thus

\begin{eqnarray}
E_n^{(1)}\,&=&\,\langle \phi_n(x) |V_{DT}(x)| \phi_n(x) \rangle\nonumber\\
&=&\,\frac{A}{\sqrt{\pi} 2^n n!}\int_{-\infty}^{+\infty} dx\,H_n^2(x) e^{-(\beta+1)x^2}, 
\label{eq:perturbative-first-order-correction}
\end{eqnarray}

where $\phi_n(x)$ are the HO functions being the solutions to the non-interacting Schr\"odinger equation with an 
HO potential and are given by

\begin{eqnarray}
&&\phi_n(x)\,=\,\frac{1}{\sqrt{2^n n! \sqrt{\pi}}} H_n(x) e^{-x^2/2}. \hskip1cm (n=1,2,3,\cdots)\nonumber\\
\label{eq:HO-func-for-perturbation-theory}
\end{eqnarray}

The second-order correction arising from a number of levels $M$ is thus

\begin{eqnarray}
E_n^{(2)}\,&=&\,\sum_{\begin{smallmatrix} k \\ k \ne n \end{smallmatrix}}^M 
\frac{| \langle\phi_n(x)| V_{DT}(x)| \phi_k(x)\rangle|^2}{E_k^{(0)}\,-\,E_n^{(0)}},\nonumber\\
&=&\,\frac{A^2}{\pi}\sum_{\begin{smallmatrix} k \\ k \ne n \end{smallmatrix}}^M 
\frac{1}{2^{n+k} n! k!(k-n)}\,\times\nonumber\\
&&\left|\displaystyle\int_{-\infty}^{+\infty} dx\,H_n(x) H_k(x) e^{-(\beta+1) x^2}\right|^2.
\label{eq:perturbative-second-order-correction}
\end{eqnarray}

The above Eqs.(\ref{eq:perturbative-first-order-correction}) and 
(\ref{eq:perturbative-second-order-correction}) can be evaluated numerically by 
Mathematica. The energy level $E_n$ of the DBEC in the HO+NGP is then the sum of 
the latter corrections plus the unperturbed energy level 

\begin{equation}
E_n^{(0)}=\left(n+\frac{1}{2}\right)
\end{equation}

such that

\begin{equation}
E_n\,=\,E_n^{(0)}\,+\,E_n^{(1)}\,+\,E_n^{(2)}.
\label{eq:perturbative-total-energy-level}
\end{equation}

Table \ref{tab:HO+NGP-matched-to-GP-energies} lists for example 
the $\langle E_{GP}\rangle_t$ of the DBEC at the resonant values of ${\cal G}_{1dd}$ for the system in 
Fig.\ft\ref{fig:plotDataEFTSignalvsGdipG1DA30B4Powx2Lx1Wext0Freq10sevdA}(A) at $\delta A=10$ with 
corresponding values $E_n$ obtained from Eqs.(\ref{eq:perturbative-first-order-correction}--\ref{eq:perturbative-total-energy-level}). 
The values of $\langle E_{GP}\rangle_t$ and $E_n$ agree very well after matching and consequently an analytical 
ELS can be deduced. 

\hs It makes sense to state that when $\langle E_{GP}\rangle_t$ matches with an $E_n$ the PR occurs. 
Therefore, a Green's function of the form 

\begin{displaymath}
G(E)\,\sim\,\frac{1}{E\,-\,E_n\,+\,i\Gamma}
\label{eq:hypthetical-Greens-function}
\end{displaymath}

could be proposed that accounts for these PRs, where $\Gamma$ is a width and 
$E=\langle E_{GP}\rangle_t$. The latter $\langle E_{GP}\rangle_t$ increase with
\Gkdd\ while $|\langle V_{eff}(0)\rangle|$ [see Sec.\ft\ref{sec:effective-potential} 
and Eq.(\ref{eq:effectivepotential-time-averaged})] drops with it. The quantum number 
$n$ rises as well indicating that the PRs shift to higher frequencies. In fact, this 
can be related to the results of the experiment of Balik \ea\ \cite{Balik:2009} who 
excited PRs via a CO${}_2$ laser with an intensity modulated via a depth $h$. They 
showed that a reduction in $h$ shifts the PRs to higher frequencies. Our findings 
above are analogous to theirs for this case.

\begin{table*}
\caption{As in Table \ref{tab:HO+NGP-matched-to-GP-energies}; but for the BOX+NGP 
in Fig.\ft\ref{fig:plotDataEFTSignalvsGdipG1D50A30B4Powx100Lx100Wext0Freq10sevdA}($A$) with the 
corrections to the energies obtained via $\phi_n(x)$ given by Eq.(\ref{eq:box-wave-function}). 
$\langle E_{GP}\rangle_t$ and $E_n$ are in units of \hw, and ${\cal G}_{1dd}$ in \aho.}
\begin{tabular}{|*6{@{\hspace{0.5cm}} c @{\hspace{0.5cm}} |}} \hline
\Gkdd\    & $\langle E_{GP}\rangle_t$ &     $E_n$    &   $n$   &   $M$   & $|\langle E_{GP} \rangle_t\,-\,E_n|$ \\ \hline
   7.0    &            8.00596        &    7.96441   &   131   &  129    &  0.04155  \\   
  78.0    &            5.44211        &    5.43156   &   109   &  101    &  0.01055  \\
 146.882  &            3.55976        &    3.56896   &    89   &   81    &  0.00920  \\
 216.331  &            3.27785        &    3.27527   &    87   &   21    &  0.00258  \\
 285.841  &            3.30829        &    3.31255   &    85   &   81    &  0.00426  \\
 356.060  &            2.23922        &    2.25459   &    73   &  103    &  0.01537  \\ \hline
\end{tabular}
\label{tab:BOX+NGP-matched-to-GP-energies}
\end{table*}

\subsection{Energy Levels in a BOX+NGP trap via Perturbation Theory}\label{sec:perturbation-theory-energy-levels-BOX}

\hs The goal of this section is the same as the prior one, but for the box potential. The 
wave function $\phi_n(x)$ for a 1D box of length $2 L_x$ is appropriately 

\begin{equation}
\phi_n(x)\,=\,\frac{1}{\sqrt{L_x}} \cos\left(\frac{n\pi x}{2 L_x}\right),\hspace{0.5cm} (n=1, 3, 5, \cdots)
\label{eq:box-wave-function}
\end{equation}

again being the solution to the box-potential Schr\"odinger equation in the noninteracting case. 
Similar to Eqs.(\ref{eq:perturbative-first-order-correction}) and 
(\ref{eq:perturbative-second-order-correction}) this yields

\begin{equation}
E_n^{(1)}\,=\,\frac{A}{L_x}\,\int_{-L_x}^{+L_x} dx\,\left[\cos\left(\frac{n\pi x}{2 L_x}\right)\right]^2 
e^{-\beta x^2}
\label{eq:first-order-correction-box}
\end{equation}

and 

\begin{eqnarray}
&&E_n^{(2)}\,=\,\frac{8 A^2}{\pi^2}\,\sum_{\begin{smallmatrix} k \\ k \ne n \end{smallmatrix}}^M 
\frac{1}{(k^2-n^2)}\times\nonumber\\
&&\left|\int_{-L_x}^{+L_x} dx\,
\cos\left(\frac{n\pi x}{2 L_x}\right) \cos\left(\frac{k\pi x}{2 L_x}\right) e^{-\beta x^2}\right|^2,
\label{eq:second-order-correction-box}
\end{eqnarray}

where the unperturbed energies are given by

\begin{equation}
E_n^{(0)}\,=\,\frac{n^2 \pi^2 }{8 L_x^2}
\end{equation}

(which are in trap units). Accordingly, Table \ref{tab:BOX+NGP-matched-to-GP-energies} lists the same
quantities as in Table \ref{tab:HO+NGP-matched-to-GP-energies}; but for the BOX+NGP. It is 
found again that perturbation theory is able to reproduce the GP energies at which the PRs 
occur. It makes sense also here to state that when $\langle E_{GP}\rangle_t$ matches $E_n$ a PR arises. 
Notably, the PRs correspond to very high energy levels $n$ because the size of the system $2L_x$ 
is quite large ($L_x=51.2$) yielding small differences in the $E_n$ and thereby a large density of states. 
However, the behavior of $\langle E_{GP}\rangle_t$ with \Gkdd\ is opposite to that of the HO$+$NGP. 
The reduction in the depth of the effective potential with \Gkdd\ causes the PR energies to drop 
shifting them to lower frequencies instead.  Thus the DDI in the case of a BOX$+$NGP cannot prevent 
the bosons from falling deeper into the NGP. But why does this happen for the box and not the HO? 
Two reasons come to mind: (1) the density in the BOX$+$NGP is lower than in the HO$+$NGP as a result 
of which the dipolar nonlinearity is weaker in the box; (2) the quantum pressure in the HO trap is 
larger than in the box. This finding, in conjunction with the one in the previous section, demonstrates 
the role of an interplay between the trapping geometry and interactions in defining the positions of 
the PRs in the spectrum of \Dkin.


\begin{table*}
\caption{As in Table \ref{tab:HO+NGP-matched-to-GP-energies}; but for the QT$+$NGP 
in Fig.\ft\ref{fig:plotDataEFTSignalvsGdipG1DA30B4Powx4Lx1Wext0Freq10sevdA}(A) with the 
corrections to the energies obtained via $\phi_n(x)$ given by Eq.(\ref{eq:WKB-wave-function}) with
the wave vectors (\ref{eq:WKB-wave-vectors}). $\langle E_{GP}\rangle_t$ and $E_n$ are in units of \hw, 
and ${\cal G}_{1dd}$ in \aho.}
\begin{tabular}{|*6{@{\hspace{0.5cm}} c @{\hspace{0.5cm}} |}} \hline
  \Gkdd\   & $\langle E_{GP}\rangle_t$ &    $E_n$    &  $n$  &  $M$  & $|\langle E_{GP} \rangle_t\,-\,E_n|$ \\ \hline
  173.701  &            23.9272        &  24.8019    &  10   &  60   &  0.87470 \\   
  245.888  &            31.7848        &  31.409     &  12   &  60   &  0.37580 \\ \hline
\end{tabular}
\label{tab:QT+NGP-matched-to-GP-energies}
\end{table*}


\subsection{Energy Levels in a QT+NGP via Perturbation Theory}\label{sec:perturbation-theory-energy-levels-QT}

\hs The energy levels of a general power-law trap 

\begin{equation}
V_{tr}(x)\,=\,\alpha |x|^{\nu}
\label{eq:general-power-law}
\end{equation}

can be obtained from WKB theory as described in advanced quantum mechanics textbooks 
\cite{Griffiths:2018} and yields (by setting $\hbar=m=1$)

\begin{equation}
E_n^{(0)}\,=\,\alpha\left[\left(n\,+\,\frac{1}{2}\right)\sqrt{\frac{\pi}{2\alpha}}
\frac{\Gamma\left(\frac{1}{\nu}\,+\,\frac{3}{2}\right)}
{\Gamma(\frac{1}{\nu}\,+\,1)}\right]^{\frac{2\nu}{\nu\,+\,2}}.
\label{eq:energy-levels-power-law-trap}
\end{equation}

The left and right turning points $a$ and $b$, respectively, are obtained from 

\begin{equation}
a=-b=-\left(\frac{E_n^{(0)}}{\alpha}\right)^{1/\nu},
\label{eq:turning-points}
\end{equation}

where in the present case $\alpha=1/2$ and $\nu=4$. The WKB wave functions are determined by the 
usual matching of the solutions at the turning points and read

\begin{equation}
\phi_n(x)\,=\,\left\{\begin{array}{cc}
\displaystyle\frac{R}{\sqrt{\kappa_n(x)}}e^{-\int_x^a \kappa_n(y) dy}; & x\ll a \\
\displaystyle\frac{2R}{\sqrt{k_n(x)}}\sin\left[\int_x^b k_n(y) dy\,+\,\frac{\pi}{4}\right]; & a<x<b \\
\displaystyle\frac{R}{\sqrt{\kappa_n(x)}} e^{-\int_b^x \kappa_n(y) dy}; & x\gg b
\end{array} \right.
\label{eq:WKB-wave-function}
\end{equation}

with

\begin{eqnarray}
&& k_n(x)\,=\,\sqrt{2[E_n^{(0)}\,-\,V_{tr}(x)]}, \nonumber\\
&& \kappa_n(x)\,=\,ik(x),
\label{eq:WKB-wave-vectors}
\end{eqnarray}

the wave vectors. $R$ is a constant determined by normalizing the
ground-state wave function at $n=0$, that is $\int_{-\infty}^{+\infty} |\phi_0(x)|^2\,=\,1$
so that

\begin{eqnarray}
&&\displaystyle \int_{-\infty}^a dx \left|\frac{R}{\sqrt{\kappa_0(x)}} e^{-\int_x^a \kappa_0(y) dy}\right|^2\,+\,\nonumber\\
&&\int_a^b dx\left|\frac{2R}{\sqrt{k_0(x)}}\sin\left[\int_x^b k_0(y) dy\,+\,\frac{\pi}{4}\right]\right|^2\,+\,\nonumber\\
&&\int_b^{+\infty} dx\left|\frac{R}{\sqrt{\kappa_0(x)}} e^{-\int_x^a \kappa_0(y) dy}\right|^2\,=\,1.\nonumber\\
\label{eq:normalization-condition-WKB}
\end{eqnarray}

The first-order correction to the unperturbed energy $E_n^{(0)}$ [Eq.(\ref{eq:energy-levels-power-law-trap})] 
via the NGP perturbation becomes 

\begin{equation}
E_n^{(1)}\,=\,4 R^2 A\,\int_a^b dx \frac{1}{k_n(x)}\sin^2\left[\int_x^b k_n(x) dx\,+\,\frac{\pi}{4}\right] e^{-\beta x^2}.
\end{equation}

It should be noted, that the wave functions in the regimes outside the QT have not been
considered in evaluating $E_n^{(1)}$ since they do not overlap with the NGP and therefore do not
make any contribution. Along the same lines, the second-order correction yields

\begin{eqnarray}
&&E_n^{(2)}\,=\,\sum_{\begin{smallmatrix} m \\ m \ne n \end{smallmatrix}}\frac{1}{E_m^{(0)}\,-\,E_n^{(0)}}\times\nonumber\\ 
&&\left| 4 R^2 A \displaystyle
\int_a^b dx \frac{1}{\sqrt{k_n(x) k_m(x)}}e^{-\beta x^2}\times\right.\nonumber\\
&&\left.\sin\left[\int_x^b k_n(y) dy\,+\,\frac{\pi}{4}\right]
\sin\left[\int_x^b k_m(y) dy\,+\,\frac{\pi}{4}\right]\right|^2, \nonumber\\
\end{eqnarray}

where $E_n^{(0)}$ is given by (\ref{eq:energy-levels-power-law-trap}).

\hs Table \ref{tab:QT+NGP-matched-to-GP-energies} shows again that $\langle E_{GP}\rangle_t$ 
at the PRs in the QT can be closely matched to the energies of second-order perturbation theory. 
It should be emphasized that well-behaved, exact analytic solutions to the Schr\"odinger equation 
with a quartic oscillator are hitherto unknown. Although, the Heun function \cite{Ronveaux:1995} 
may provide a solution within a restricted range in the neighborhood of the trap center, outside 
the latter it diverges to very large values. We have found that even a numerical solution by 
Mathematica yields divergent solutions away from the center (not shown here).

\subsection{Transition Probabilities}

\subsubsection{Harmonic Trap}

\hs In what follows, the transition probabilities at which PRs occur in an HO+NGP trap are evaluated. Basing on the assumption 
that the NGP and DDI are perturbations inducing transitions between the different states in the traps, the probabilities for
these can be computed according to time-dependent perturbation theory. In this
regard, the probability for a transfer from some state $m$ to state $n$ is from standard quantum
mechanics textbooks given by

\begin{equation}
P(n,m,t)\,=\,|c_{n,m}(t)|^2,
\label{eq:P(n,m,t)}
\end{equation}

where as usual

\begin{equation}
c_{n,m}(t)\,=\,\frac{\lambda}{i\hbar} \int_0^t dt^\prime e^{i(E_m^{(0)}-E_n^{(0)})t^\prime/\hbar} 
\langle\phi_m(x)|V(t^\prime)|\phi_n(x)\rangle
\label{eq:coefficient-cn(t)-from-time-dependent-PertTh}
\end{equation}

with $V(t^\prime)$ being a time-dependent perturbation. Considering first $\phi_n(x)$ to be 
the HO states given by (\ref{eq:HO-func-for-perturbation-theory}) and $V(t^\prime)$ the 
time-dependent NGP (\ref{eq:NGP-with-osc-depth}), then it is easy to show that the probability 
introduced by the NGP is (setting $\hbar=1$)

\begin{eqnarray}
&&P^{DT}(n,m,t)\,=\,\left|\frac{\lambda}{i} \frac{1}{\sqrt{\pi\,2^{n+m}\,n!\,m!}}\times\right.\nonumber\\ 
&&\left.\left\{\int_{-\infty}^{+\infty} H_m(x) H_n(x) e^{-(\beta+1) x^2} dx\right\}\times\right.\nonumber\\
&&\left.\Bigg[A F_0(m,n,t)\,+\,\delta A (F_{+\Omega}(m,n,t)\,+\,F_{-\Omega}(m,n,t)\Bigg]\right|^2,\nonumber\\
\label{eq:probability-of-DT-induced-transitions}
\end{eqnarray}

where

\begin{eqnarray}
F_0(m,n,t)\,&=&\,\frac{e^{i(m-n)t}\,-\,1}{i(m-n)}, \nonumber\\
F_{+\Omega}(m,n,t)\,&=&\,\frac{e^{i(m-n+\Omega)t}\,-\,1}{2i(m-n+\Omega)}, \nonumber\\
F_{-\Omega}(m,n,t)\,&=&\,\frac{e^{i(m-n-\Omega)t}\,-\,1}{2i(m-n-\Omega)}.
\end{eqnarray}

On the other hand, to compute the transition probability because of 
the DDI, the Fourier transform (FT) of the 1D dipolar potential $V_{1D}(x-x^{\prime})$, 
such as the one presented by Sinha and Santos \cite{Sinha:2007}, needs to be applied to avoid 
the singularity in the DDI. This FT reads for a quasi 1D Bose gas

\begin{equation}
\tilde{V}_{1D}(k)\,=\,\frac{4\alpha d^2}{\ell^2}\Bigg[1\,-\,\sigma e^\sigma \Gamma(0,\sigma)\Bigg],
\label{eq:SinhasFourierTransformoftheDDIpotential}
\end{equation}

were $\Gamma(0,\sigma)$ is the incomplete gamma function, and $\sigma=k^2\ell^2/4$. Here 
$\ell=\sqrt{\hbar/(\mu \bar{\omega})}$ is a length scale with $\mu=m/2$ the reduced mass. The factor $\alpha$ has
values between $-1/2$ ($\phi=\pi/2$) and $1$ ($\phi=0$) where $\phi$ is the angle between
the dipoles and the longitudinal axis of the trap. In the present case $\ell=d_{\rho}$ 
is the width of the wave function in the transverse direction. Appendix \ref{sec:V1Dkunits} 
shows how to cast (\ref{eq:SinhasFourierTransformoftheDDIpotential}) in trap units, that is 
$\tilde{V}_{1D}\rightarrow\tilde{V}_{1D}/(\hbar\omega)$.

\hs Following Muruganandam \cite{Muruganandam:2012}, the spatial integral required in 
Eq.(\ref{eq:coefficient-cn(t)-from-time-dependent-PertTh}) for the transition between 
states $n$ and $m$, namely 

\begin{eqnarray}
&&\langle\phi_n(x)|V(t^\prime)|\phi_m(x)\rangle\,\rightarrow\,\nonumber\\
&&\frac{N}{2} \int_{-\infty}^{+\infty} dx \int_{-\infty}^{+\infty} 
dx^\prime \phi_n^\star(x) \phi_m(x^\prime) V_{1D}(x-x^\prime),\nonumber\\
\end{eqnarray}

can be rewritten using the FT [Eq.(\ref{eq:SinhasFourierTransformoftheDDIpotential})] with its strength given 
by (\ref{eq:coefficient-of-Vdd(k)-final-form}) and the FT of the HO functions $\widetilde{\phi}_n(k)$ \cite{Sakhel:2018} where 

\begin{equation}
\widetilde{\phi}_n(k)\,=\,{\rm FT}[\phi_{n}(x)]\,=\,\frac{1}{\sqrt{\sqrt{\pi} 2^n n!}} (-i)^n H_n(k) e^{-k^2/2}.
\end{equation}

This then yields that 

\begin{eqnarray}
&&\langle\phi_n(x)|V(t^\prime)|\phi_m(x)\rangle\,\rightarrow\,\nonumber\\
&&\frac{N}{2}\frac{1}{2\pi} \int_{-\infty}^{+\infty} dk\,\widetilde{\phi}_n^\star(k) 
\widetilde{\phi} _m(-k) \tilde{V}_{1D}(k).
\end{eqnarray}

As such, the probability for the DDI reads then

\begin{eqnarray}
&&P^{DDI}(n,m,t)\,=\,\lambda^2\left|\frac{{\cal G}_{1dd} (-i)^{n+m}}{\sqrt{\pi\,2^{n+m}\,n!\,m!}} \times\right.\nonumber\\
&&\left.\int_{-\infty}^{+\infty} H_n(k) H_m(-k) e^{-k^2}\times\right.\nonumber\\
&&\left.\left[1-\frac{k^2 d^2_{\rho}}{4} e^{k^2 d^2_{\rho}/4} \Gamma(0,k^2 d^2_\rho/4)\right] F_0(m,n,t)\right|^2.\nonumber\\
\label{eq:probability-of-DDI-induced-transitions}
\end{eqnarray}

The above two probabilities $P^{DT}(n,m,t)$ [Eq.(\ref{eq:probability-of-DT-induced-transitions})] and $P^{DDI}(n,m,t)$ 
can be easily evaluated numerically by Mathematica for the integer values of $m$ and $n$. These two probabilities are 
dependent on each other and therefore the total probability is 

\begin{equation}
{\cal P}(n,m,t)\,=\,P^{DT}(n,m,t) P^{DDI}(n,m,t)
\label{eq:total-transition-probability}.
\end{equation}

\hs Table \ref{tab:transition-probabilities-HO-NGP} displays $P(n,m,t)$ (in units of 
arbitrary $\lambda^4$) for the most probable transitions to the states $n$ matched in 
Table \ref{tab:HO+NGP-matched-to-GP-energies} for a length of one driving cycle $t=0.1$. 
In this, it is assumed that $\lambda\ll 1$. It should be noted, that a significant number 
of $m\leftrightarrow n$ transitions turned out to be prohibited with identically zero 
values for $P(n,m,t)$. This explains the absence of PRs for values of \Gkdd\ with a 
time-averaged GP that nevertheless can be matched to one of the energy levels 
given by perturbation theory in Sec.\ft\ref{sec:perturbation-theory-energy-levels-HO}.
Thus there are two conditions for PRs to occur: (1) its energy should match one of the
trap levels and (2) the transition to this level should have a high probability.

\begin{table}
\caption{Most probable transitions to the states $n$ at which the GP energies are 
matched in Table \ref{tab:HO+NGP-matched-to-GP-energies} for the HO+NGP trap. The transition 
probability is computed by Eqs.(\ref{eq:probability-of-DT-induced-transitions}), 
(\ref{eq:probability-of-DDI-induced-transitions}),  and (\ref{eq:total-transition-probability}). 
${\cal G}_{1dd}$ is in units of $a_{ho}$.}
\begin{tabular}{|*4{@{\hspace{0.5cm}} c @{\hspace{0.5cm}} |}} \hline
 \Gkdd\   &  $m$  &  $n$  &   ${\cal P}(n,m,t)/\lambda^4$ \\ \hline
 29.9297  &   2   &   4   &  0.0261 \\
 67.8894  &   5   &   7   &  0.0511 \\
 120.4490 &   8   &  10   &  0.1216 \\
 191.9880 &  11   &  13   &  0.2210 \\ 
 292.7270 &  16   &  18   &  0.3430 \\ \hline\hline
\end{tabular}
\label{tab:transition-probabilities-HO-NGP}
\end{table}

\subsubsection{Box Potential}

\hs As in the previous section, the goal is to find the most probable transitions
at which the PRs arise in the BOX$+$NGP trap. In this case, 

\begin{eqnarray}
&&P^{DT}(n,m,t)\,=\,\left|\frac{\lambda}{i}\frac{1}{L_x}\times\right.\nonumber\\
&&\left.\left\{\int_{-L_x}^{+L_x} \cos\left(\frac{m\pi x}{2L_x}\right) 
\cos\left(\frac{n\pi x}{2L_x}\right) e^{-\beta x^2} dx\right\}\times\right.\nonumber\\
&&\left.\Bigg[A F_0(m,n,t)\,+\,\delta A (F_{+\Omega}(m,n,t)\,+\,F_{-\Omega}(m,n,t))\Bigg]\right|^2\nonumber\\
\label{eq:prob-of-DT-induced-transitions-in-a-box}
\end{eqnarray}

and 

\begin{eqnarray}
&&P^{DDI}(n,m,t)\,=\,\left|\frac{\lambda}{i} 4 L_x^2 {\cal G}_{1dd} \times\right.\nonumber\\ 
&&\left.\int_{-K_0}^{+K_0}\,\left[\frac{\sin\left(\displaystyle kL_x\,-\,\frac{n\pi}{2}\right)}{2kL_x\,-\,n\pi}\,+\,
\frac{\sin\left(\displaystyle kL_x\,+\,\frac{n\pi}{2}\right)}{2kL_x\,+\,n\pi}\right] \right.\times\nonumber\\
&& \left.\left[\frac{\sin\left(\displaystyle kL_x\,-\,\frac{m\pi}{2}\right)}{2kL_x\,-\,m\pi}\,+\,
\frac{\sin\left(\displaystyle kL_x\,+\,\frac{m\pi}{2}\right)}{2kL_x\,+\,m\pi}\right]\times\right.\nonumber\\
&&\left.\left[1-\frac{k^2 d^2_{\rho}}{4} e^{k^2 d^2_{\rho}/4} 
\Gamma[0,k^2 d^2_\rho/4]\right] F_0(m,n,t)\right|^2,
\label{eq:prob-of-DDI-induced-transitions-in-a-box}
\end{eqnarray}

where the FT of $\cos\left(n\pi x/(2L_x)\right)$ inside the box is

\begin{eqnarray}
&&FT[\cos\left(n\pi x/(2L_x)\right)]\,=\,\nonumber\\
&& 2L_x\left[\frac{\sin\left(\displaystyle kL_x\,-\,
\frac{n\pi}{2}\right)}{2kL_x\,-\,n\pi}\,+\,
\frac{\sin\left(\displaystyle kL_x\,+\,\frac{n\pi}{2}\right)}{2kL_x\,+\,n\pi}\right].\nonumber\\
\label{eq:Fourier-Transform-of-Cosine-Function-in-Box}
\end{eqnarray}

In (\ref{eq:prob-of-DDI-induced-transitions-in-a-box}), $K_0$ is a cutoff momentum 
introduced to simplify the integration, but which would still give the same result as 
going to infinite values. Table \ref{tab:transition-probabilities-BOX+NGP} is as in 
Table \ref{tab:transition-probabilities-HO-NGP}, but for the box. The most probable 
transitions are found to come from relatively high states $m$ towards $n$ matched in 
Table \ref{tab:BOX+NGP-matched-to-GP-energies}.

\hs Finally, it must be noted that because of the divergent wavefunction of the QT$+$NGP 
already mentioned in Sec.\ft\ref{sec:origins-of-PRs}E, an analytical evaluation of the 
$P^{DT}$ and $P^{DDI}$ is currently extremely difficult if not impossible, and is therefore 
left for the future.

\begin{table}
\caption{As in Table \ref{tab:transition-probabilities-HO-NGP}; but for the BOX+NGP trap. 
${\cal G}_{1dd}$ is in units of $a_{ho}$.}
\begin{tabular}{|*4{@{\hspace{0.5cm}} c @{\hspace{0.5cm}} |}} \hline
 \Gkdd\   &  $m$  &  $n$   &   ${\cal P}(n,m,t)/\lambda^4$ \\ \hline
   7.0    &  129  &   131  &   0.0048   \\   
  78.0    &  101  &   109  &   0.2623   \\
 146.882  &   61  &    89  &   1.1894   \\
 216.331  &   65  &    87  &   2.8498   \\
 285.841  &   81  &    85  &   3.5049   \\
 356.060  &   71  &    73  &   71.203   \\ \hline\hline
\end{tabular}
\label{tab:transition-probabilities-BOX+NGP}
\end{table}


\section{Summary and Conclusions}\label{sec:summary-and-conclusions}

\hs In summary then, we have reported mean-field parametric resonances (PRs) in a one-dimensional (1D) 
dipolar Bose-Einstein condensate (DBEC) excited by a negative Gaussian potential (NGP) with a periodically 
oscillating depth. The PRs have been detected by the signal energy \cite{Oppenheim:2015}, a quantity that 
closely resembles the time-average of the square of the energy (in this work the kinetic energy). It is 
similar to the root-mean-squared (RMS) value of an oscillating electrical signal arising from an alternating 
voltage or current source. The latter NGP was for modelling a red laser light source with oscillating 
intensity in a manner similar to Refs.\cite{Balik:2009,Clark:2015}. 

\hs By a matching of the PR energies to energy levels computed by time-independent perturbation theory 
we were able to characterize the energy-level structure (ELS) of a DBEC in a complex trap, i.e., a trap
to which an NGP has been added. Within the purpose of the latter characterization, a few different traps 
have been applied to test the effect of confining geometry and its interplay with dipole-dipole interactions 
(DDI) on the PRs. It turns out the DDI play  an important role in defining the amplitude of the PRs and their 
energies. The DDI reduces the depth of the effectve mean-field potential, thereby controlling the occupancy 
of the NGP, and in turn the positions and strengths of the PRs.

\hs The key feature of these PRs is that their properties are sensitive to the ELS of 
the confining potential, i.e., external trap+NGP. The trapping geometry determines the wavefunction that 
describes the system, and in turn the wavefunction in conjunction with the NGP and DDI determine the transition 
probabilities between different quantum states of the confinement. The PRs correspond to the most probable 
transitions determined from time-dependent perturbation theory.  Nevertheless, it is rather surprising that 
these PRs correspond only to specific values of the principal quantum numbers of high transitional probabilities, 
say $m$ and $n$, and that they do not arise at other values ($m$,$n$) of comparable, or slightly lower, 
probabilities. One might even begin to assume the presence of magic numbers for $n$ and $m$, although this 
assertion needs to be proven.

\hs Most importantly, DBEC PRs could be astonishingly produced by the Lagrangian variational method (LVM) 
\cite{Sakhel:2017} that approximates the wavefunctions by a simple Gaussian Ansatz. However, it was not 
possible to use LVM for the box and quartic trap because the Gaussian Ansatz had been originally 
designed for the harmonic trapping case. The LVM PRs are a further manifestation of the fact that 
these resonances are an inherent feature present in the DBECs awakened by external driving agents.

\hs The present work can also be somewhat related to a previous study on resonances and dynamical 
fragmentation in a stirred BEC \cite{Tsatsos:2015} in which a series of PRs 
in the total energy arises as the rotational frequency is increased. In this regard, the authors 
conclude that fragmentation of the gas appears simultaneously with the resonant absorption of 
energy and angular momentum from the external excitation agent. Whereas their PRs are associated
with fragmentation of the BEC, in the present mean-field GP formulation there is no fragmentation. 
The PRs of the current work arise because of resonant absorption of momentum from the dynamic NGP. 
This momentum arises from the time-varying force Eq.(\ref{eq:driving-force}) which is the gradient 
of the NGP that acts along the length of the BEC. 

\hs The previous results in Secs.\ft\ref{sec:results}$A$-$C$ are also a manifestation for the role of a mean-field 
dipolar nonlinearity in defining the properties of the PRs and $-$via its interplay with the trapping geometry$-$ 
also their spacings $\Delta{\cal G}_{1dd}$. One question that arises is if similar results could be obtained 
by simulations of the same systems using the many-body Schr\"odinger equation with DDI. In this case, the DDI do not 
arise in the form of a nonlinearity and it would be interesting to make comparisons between the many-body effects of the 
DDI and that of the DDI nonlinearity.

\section*{Acknowledgement}

\hs The calculations were performed on the PARADOX-IV supercomputing facility at the Scientic Computing 
Laboratory of the Institute of Physics Belgrade, supported in part by the Ministry of Education, Science 
and Technological Development of the Republic of Serbia under project No. ON171017. We thank William 
J Mullin (UMASS, Amherst USA) for insightful comments on an earlier version of this manuscript that 
helped us to improve it substantially. The authors declare that there are no conflicts of interest.

\appendix

\section{Numerics}\label{sec:numerics}

\hs The TDGPE [Eq.(\ref{eq:one-dimensional-TDGPE})] is solved 
numerically via the split-step Crank-Nicolson (CN) method \cite{Kumar:2015,Muruganandam:2009} in real time 
for the HO ($p=2$), QT ($p=4$), and box potential traps. For the HO and QT, it is propagated along a grid of $N_x=6000$ 
pixels of size $\Delta x=0.01$ with $L_x=1$ in Eq.(\ref{eq:one-dimensional-external-power-law-trap}). 
The time step chosen is $\Delta t\,=\,1\times 10^{-4}$, the number of time steps in the transient 
run is set to $N_{pas}\,=\,10^6$, and the same in the final run $N_{run}\,=\,10^6$. For the box potential 
$N_x=2048$, $\Delta x=0.05$, $L_x=51.2$, $\Delta t=0.00025$, $N_{pas}=200000$, and $N_{run}=200000$. 
In all cases, the system is initialized with a stationary NGP of 
depth $A=-30$ via the imaginary-time CN method for a number of $N_{stp}=2\times 10^5$ time steps, and then 
taken through $N_{pas}=20000$ steps in the transient, and ending with $N_{run}=20000$ steps in the final run. 
The time step $\Delta t$ is the same as in the corresponding real-time simulations. In the CN codes that have 
been used, the final run (NRUN) just complements the transient one (NPAS). The latter could be the stage during 
which the BEC is allowed to evolve and relax to a stable state. The final run could then be the stage where 
the BEC is optionally excited by an external agent in order to examine its ensuing dynamics. The authors of 
the code originally separated the transient and final runs for organizational purposes. This is in order to 
have data files in the relaxation stage of the BEC towards a stable state separate from the one where this 
stable BEC is suddenly excited by an external driving force. In contrast, the present work considers a BEC that is 
continuously excited in both transient and final runs, so there is no difference between both of them. 

\hs Sets of runs were performed each at a fixed value of ${\cal G}_{1D}$. For each ${\cal G}_{1D}$,
the system was scanned along a range of \Gkdd\ from 0 to 400 in steps of 2. It should be emphasized 
that for each \Gkdd\ there was a run. This large number of runs has been conducted in the form of 
parallel array jobs on the high performance computational cluster of the Scientific Computing Laboratory of 
the Institute of Physics in Belgrade, Serbia. Each simulation took about five days to finish. In essence, 
this has been a heavily computational project.

\section{Units of $V_{1D}(k)$}\label{sec:V1Dkunits}

\hs Beginning with the dipole moment $d$ where \cite{Kumar:2015} 

\begin{equation}
d^2\,=\,\frac{\mu_0\tilde{\mu}^2}{4\pi}
\label{eq:d2}
\end{equation}

and the dipolar scattering length 

\begin{equation}
a_{dd}\,=\,\frac{\mu_0\tilde{\mu}^2 m}{12 \pi \hbar^2},
\label{eq:add}
\end{equation}

$\tilde{\mu}$ being the magnetic dipole moment and $\mu_0$ the magnetic 
permeability, we substitute (\ref{eq:add}) into (\ref{eq:d2}) to eliminate
$\tilde{\mu}$ so that $d^2\,=\,3\hbar^2 a_{dd}/m$. Then the rescaling 
$d^2\,\rightarrow\,d^2/(\hbar\omega)$ yields

\begin{equation}
\tilde{d}^2\,=\,\frac{d^2}{\hbar\omega}\,=\,\frac{3\hbar^2 a_{dd}}{m \hbar\omega}\,=\,3 \tilde{a}_{dd} a_{ho}^3,
\end{equation}

where $\tilde{a}_{dd}\,=\,a_{dd}/a_{ho}$. Hence, the coefficient of Eq.(\ref{eq:SinhasFourierTransformoftheDDIpotential})
becomes in trap units

\begin{equation}
\frac{4\alpha \tilde{d}^2}{d_\rho^2}\,=\,\frac{4\alpha}{\tilde{d}_{\rho}^2} 3 \tilde{a}_{dd} a_{ho},
\label{eq:coefficient-of-Vdd(k)}
\end{equation}

with $\tilde{d}_{\rho}\,=\,d_{\rho}/a_{ho}$. Using Eq.(\ref{eq:definition-of-G1D}) for ${\cal G}_{1dd}$, 
one eventually gets that the coefficient with $\alpha=-1/2$ is 

\begin{equation}
\frac{4\alpha d^2}{\ell^2}\,\rightarrow\,\frac{4\pi {\cal G}_{1dd} a_{ho}}{N}.
\label{eq:coefficient-of-Vdd(k)-final-form}
\end{equation}

\bibliography{DipolarBECexcitedbyaDimple8Jan2021,red-laser,LVM}

\begin{thebibliography}{102}
\expandafter\ifx\csname natexlab\endcsname\relax\def\natexlab#1{#1}\fi
\expandafter\ifx\csname bibnamefont\endcsname\relax
  \def\bibnamefont#1{#1}\fi
\expandafter\ifx\csname bibfnamefont\endcsname\relax
  \def\bibfnamefont#1{#1}\fi
\expandafter\ifx\csname citenamefont\endcsname\relax
  \def\citenamefont#1{#1}\fi
\expandafter\ifx\csname url\endcsname\relax
  \def\url#1{\texttt{#1}}\fi
\expandafter\ifx\csname urlprefix\endcsname\relax\def\urlprefix{URL }\fi
\providecommand{\bibinfo}[2]{#2}
\providecommand{\eprint}[2][]{\url{#2}}

\bibitem[{\citenamefont{Balik et~al.}(2009)\citenamefont{Balik, Win, and
  Havey}}]{Balik:2009}
\bibinfo{author}{\bibfnamefont{S.}~\bibnamefont{Balik}},
  \bibinfo{author}{\bibfnamefont{A.~L.} \bibnamefont{Win}}, \bibnamefont{and}
  \bibinfo{author}{\bibfnamefont{M.~D.} \bibnamefont{Havey}},
  \bibinfo{journal}{Phys. Rev. A} \textbf{\bibinfo{volume}{80}},
  \bibinfo{pages}{023404} (\bibinfo{year}{2009}),
  \urlprefix\url{https://link.aps.org/doi/10.1103/PhysRevA.80.023404}.

\bibitem[{\citenamefont{Sakhel and Sakhel}(2020)}]{Sakhel:2020}
\bibinfo{author}{\bibfnamefont{R.~R.} \bibnamefont{Sakhel}} \bibnamefont{and}
  \bibinfo{author}{\bibfnamefont{A.~R.} \bibnamefont{Sakhel}},
  \bibinfo{journal}{J. Phys: Condens. Matter} \textbf{\bibinfo{volume}{32}},
  \bibinfo{pages}{315401} (\bibinfo{year}{2020}),
  \urlprefix\url{https://iopscience.iop.org/article/10.1088/1361-648X/ab7f06}.

\bibitem[{\citenamefont{Nguyen et~al.}(2019)\citenamefont{Nguyen, Tsatsos, Luo,
  Lode, Telles, Bagnato, and Hulet}}]{Nguyen:2019}
\bibinfo{author}{\bibfnamefont{J.~H.~V.} \bibnamefont{Nguyen}},
  \bibinfo{author}{\bibfnamefont{M.~C.} \bibnamefont{Tsatsos}},
  \bibinfo{author}{\bibfnamefont{D.}~\bibnamefont{Luo}},
  \bibinfo{author}{\bibfnamefont{A.~U.~J.} \bibnamefont{Lode}},
  \bibinfo{author}{\bibfnamefont{G.~D.} \bibnamefont{Telles}},
  \bibinfo{author}{\bibfnamefont{V.~S.} \bibnamefont{Bagnato}},
  \bibnamefont{and} \bibinfo{author}{\bibfnamefont{R.~G.} \bibnamefont{Hulet}},
  \bibinfo{journal}{Phys. Rev. X} \textbf{\bibinfo{volume}{9}},
  \bibinfo{pages}{011052} (\bibinfo{year}{2019}),
  \urlprefix\url{https://link.aps.org/doi/10.1103/PhysRevX.9.011052}.

\bibitem[{\citenamefont{Chen et~al.}(2019)\citenamefont{Chen, Shibata, Eto,
  Hirano, and Saito}}]{Chen:2019}
\bibinfo{author}{\bibfnamefont{T.}~\bibnamefont{Chen}},
  \bibinfo{author}{\bibfnamefont{K.}~\bibnamefont{Shibata}},
  \bibinfo{author}{\bibfnamefont{Y.}~\bibnamefont{Eto}},
  \bibinfo{author}{\bibfnamefont{T.}~\bibnamefont{Hirano}}, \bibnamefont{and}
  \bibinfo{author}{\bibfnamefont{H.}~\bibnamefont{Saito}},
  \bibinfo{journal}{Phys. Rev. A} \textbf{\bibinfo{volume}{100}},
  \bibinfo{pages}{063610} (\bibinfo{year}{2019}),
  \urlprefix\url{https://link.aps.org/doi/10.1103/PhysRevA.100.063610}.

\bibitem[{\citenamefont{Zhu et~al.}(2019)\citenamefont{Zhu, Yi, Guo, and
  Zhou}}]{Zhu:2019}
\bibinfo{author}{\bibfnamefont{C.-X.} \bibnamefont{Zhu}},
  \bibinfo{author}{\bibfnamefont{W.}~\bibnamefont{Yi}},
  \bibinfo{author}{\bibfnamefont{G.-C.} \bibnamefont{Guo}}, \bibnamefont{and}
  \bibinfo{author}{\bibfnamefont{Z.-W.} \bibnamefont{Zhou}},
  \bibinfo{journal}{Phys. Rev. A} \textbf{\bibinfo{volume}{99}},
  \bibinfo{pages}{023619} (\bibinfo{year}{2019}),
  \urlprefix\url{https://link.aps.org/doi/10.1103/PhysRevA.99.023619}.

\bibitem[{\citenamefont{Li et~al.}(2019)\citenamefont{Li, Jiang, Lan, Zhang,
  and Zhou}}]{ZhengChun:2019}
\bibinfo{author}{\bibfnamefont{Z.-C.} \bibnamefont{Li}},
  \bibinfo{author}{\bibfnamefont{Q.-H.} \bibnamefont{Jiang}},
  \bibinfo{author}{\bibfnamefont{Z.}~\bibnamefont{Lan}},
  \bibinfo{author}{\bibfnamefont{W.}~\bibnamefont{Zhang}}, \bibnamefont{and}
  \bibinfo{author}{\bibfnamefont{L.}~\bibnamefont{Zhou}},
  \bibinfo{journal}{Phys. Rev. A} \textbf{\bibinfo{volume}{100}},
  \bibinfo{pages}{033617} (\bibinfo{year}{2019}),
  \urlprefix\url{https://link.aps.org/doi/10.1103/PhysRevA.100.033617}.

\bibitem[{\citenamefont{Sakhel and Sakhel}(2018)}]{Sakhel:2018}
\bibinfo{author}{\bibfnamefont{A.~R.} \bibnamefont{Sakhel}} \bibnamefont{and}
  \bibinfo{author}{\bibfnamefont{R.~R.} \bibnamefont{Sakhel}},
  \bibinfo{journal}{Journal of Low Temperature Physics}
  \textbf{\bibinfo{volume}{190}}, \bibinfo{pages}{120} (\bibinfo{year}{2018}),
  ISSN \bibinfo{issn}{1573-7357},
  \urlprefix\url{https://doi.org/10.1007/s10909-017-1826-7}.

\bibitem[{\citenamefont{Molignini et~al.}(2018)\citenamefont{Molignini,
  Papariello, Lode, and Chitra}}]{Molignini:2018}
\bibinfo{author}{\bibfnamefont{P.}~\bibnamefont{Molignini}},
  \bibinfo{author}{\bibfnamefont{L.}~\bibnamefont{Papariello}},
  \bibinfo{author}{\bibfnamefont{A.~U.~J.} \bibnamefont{Lode}},
  \bibnamefont{and} \bibinfo{author}{\bibfnamefont{R.}~\bibnamefont{Chitra}},
  \bibinfo{journal}{Phys. Rev. A} \textbf{\bibinfo{volume}{98}},
  \bibinfo{pages}{053620} (\bibinfo{year}{2018}),
  \urlprefix\url{https://link.aps.org/doi/10.1103/PhysRevA.98.053620}.

\bibitem[{\citenamefont{Robertson et~al.}(2018)\citenamefont{Robertson, Michel,
  and Parentani}}]{Robertson:2018}
\bibinfo{author}{\bibfnamefont{S.}~\bibnamefont{Robertson}},
  \bibinfo{author}{\bibfnamefont{F.}~\bibnamefont{Michel}}, \bibnamefont{and}
  \bibinfo{author}{\bibfnamefont{R.}~\bibnamefont{Parentani}},
  \bibinfo{journal}{Phys. Rev. D} \textbf{\bibinfo{volume}{98}},
  \bibinfo{pages}{056003} (\bibinfo{year}{2018}),
  \urlprefix\url{https://link.aps.org/doi/10.1103/PhysRevD.98.056003}.

\bibitem[{\citenamefont{{S. Lellouch, M. Bukov, E. Demler, and N.
  Goldman}}(2017)}]{Lellouch:2017}
\bibinfo{author}{\bibnamefont{{S. Lellouch, M. Bukov, E. Demler, and N.
  Goldman}}}, \bibinfo{journal}{Phys. Rev. X} \textbf{\bibinfo{volume}{7}},
  \bibinfo{pages}{021015} (\bibinfo{year}{2017}).

\bibitem[{\citenamefont{Vidanovi{\'{c}}
  et~al.}(2012)\citenamefont{Vidanovi{\'{c}}, Al-Jibbouri, Bala{\v{z}}, and
  Pelster}}]{Vidanovic:2012}
\bibinfo{author}{\bibfnamefont{I.}~\bibnamefont{Vidanovi{\'{c}}}},
  \bibinfo{author}{\bibfnamefont{H.}~\bibnamefont{Al-Jibbouri}},
  \bibinfo{author}{\bibfnamefont{A.}~\bibnamefont{Bala{\v{z}}}},
  \bibnamefont{and} \bibinfo{author}{\bibfnamefont{A.}~\bibnamefont{Pelster}},
  \bibinfo{journal}{Physica Scripta} \textbf{\bibinfo{volume}{T149}},
  \bibinfo{pages}{014003} (\bibinfo{year}{2012}),
  \urlprefix\url{https://doi.org/10.1088%2F0031-8949%2F2012%2Ft149%2F014003}.

\bibitem[{\citenamefont{Posazhennikova
  et~al.}(2016)\citenamefont{Posazhennikova, Trujillo-Martinez, and
  Kroha}}]{Posazhennikova:2016}
\bibinfo{author}{\bibfnamefont{A.}~\bibnamefont{Posazhennikova}},
  \bibinfo{author}{\bibfnamefont{M.}~\bibnamefont{Trujillo-Martinez}},
  \bibnamefont{and} \bibinfo{author}{\bibfnamefont{J.}~\bibnamefont{Kroha}},
  \bibinfo{journal}{Phys. Rev. Lett.} \textbf{\bibinfo{volume}{116}},
  \bibinfo{pages}{225304} (\bibinfo{year}{2016}),
  \urlprefix\url{https://link.aps.org/doi/10.1103/PhysRevLett.116.225304}.

\bibitem[{\citenamefont{Kobyakov et~al.}(2012)\citenamefont{Kobyakov, Bychkov,
  Lundh, Bezett, and Marklund}}]{Kobyakov:2012}
\bibinfo{author}{\bibfnamefont{D.}~\bibnamefont{Kobyakov}},
  \bibinfo{author}{\bibfnamefont{V.}~\bibnamefont{Bychkov}},
  \bibinfo{author}{\bibfnamefont{E.}~\bibnamefont{Lundh}},
  \bibinfo{author}{\bibfnamefont{A.}~\bibnamefont{Bezett}}, \bibnamefont{and}
  \bibinfo{author}{\bibfnamefont{M.}~\bibnamefont{Marklund}},
  \bibinfo{journal}{Phys. Rev. A} \textbf{\bibinfo{volume}{86}},
  \bibinfo{pages}{023614} (\bibinfo{year}{2012}),
  \urlprefix\url{https://link.aps.org/doi/10.1103/PhysRevA.86.023614}.

\bibitem[{\citenamefont{Xue et~al.}(2008)\citenamefont{Xue, Li, Zhang, and
  Peng}}]{Xue:2008}
\bibinfo{author}{\bibfnamefont{J.-K.} \bibnamefont{Xue}},
  \bibinfo{author}{\bibfnamefont{G.-Q.} \bibnamefont{Li}},
  \bibinfo{author}{\bibfnamefont{A.-X.} \bibnamefont{Zhang}}, \bibnamefont{and}
  \bibinfo{author}{\bibfnamefont{P.}~\bibnamefont{Peng}},
  \bibinfo{journal}{Phys. Rev. E} \textbf{\bibinfo{volume}{77}},
  \bibinfo{pages}{016606} (\bibinfo{year}{2008}),
  \urlprefix\url{https://link.aps.org/doi/10.1103/PhysRevE.77.016606}.

\bibitem[{\citenamefont{Engels et~al.}(2007)\citenamefont{Engels, Atherton, and
  Hoefer}}]{Engels:2007a}
\bibinfo{author}{\bibfnamefont{P.}~\bibnamefont{Engels}},
  \bibinfo{author}{\bibfnamefont{C.}~\bibnamefont{Atherton}}, \bibnamefont{and}
  \bibinfo{author}{\bibfnamefont{M.~A.} \bibnamefont{Hoefer}},
  \bibinfo{journal}{Phys. Rev. Lett.} \textbf{\bibinfo{volume}{98}},
  \bibinfo{pages}{095301} (\bibinfo{year}{2007}),
  \urlprefix\url{https://link.aps.org/doi/10.1103/PhysRevLett.98.095301}.

\bibitem[{\citenamefont{Nicolin et~al.}(2007)\citenamefont{Nicolin,
  Carretero-Gonz\'alez, and Kevrekidis}}]{Nicolin:2007}
\bibinfo{author}{\bibfnamefont{A.~I.} \bibnamefont{Nicolin}},
  \bibinfo{author}{\bibfnamefont{R.}~\bibnamefont{Carretero-Gonz\'alez}},
  \bibnamefont{and} \bibinfo{author}{\bibfnamefont{P.~G.}
  \bibnamefont{Kevrekidis}}, \bibinfo{journal}{Phys. Rev. A}
  \textbf{\bibinfo{volume}{76}}, \bibinfo{pages}{063609}
  (\bibinfo{year}{2007}),
  \urlprefix\url{https://link.aps.org/doi/10.1103/PhysRevA.76.063609}.

\bibitem[{\citenamefont{Kr\"amer et~al.}(2005)\citenamefont{Kr\"amer, Tozzo,
  and Dalfovo}}]{Kraemer:2005}
\bibinfo{author}{\bibfnamefont{M.}~\bibnamefont{Kr\"amer}},
  \bibinfo{author}{\bibfnamefont{C.}~\bibnamefont{Tozzo}}, \bibnamefont{and}
  \bibinfo{author}{\bibfnamefont{F.}~\bibnamefont{Dalfovo}},
  \bibinfo{journal}{Phys. Rev. A} \textbf{\bibinfo{volume}{71}},
  \bibinfo{pages}{061602} (\bibinfo{year}{2005}),
  \urlprefix\url{https://link.aps.org/doi/10.1103/PhysRevA.71.061602}.

\bibitem[{\citenamefont{Tozzo et~al.}(2005)\citenamefont{Tozzo, Kr\"amer, and
  Dalfovo}}]{Tozzo:2005}
\bibinfo{author}{\bibfnamefont{C.}~\bibnamefont{Tozzo}},
  \bibinfo{author}{\bibfnamefont{M.}~\bibnamefont{Kr\"amer}}, \bibnamefont{and}
  \bibinfo{author}{\bibfnamefont{F.}~\bibnamefont{Dalfovo}},
  \bibinfo{journal}{Phys. Rev. A} \textbf{\bibinfo{volume}{72}},
  \bibinfo{pages}{023613} (\bibinfo{year}{2005}),
  \urlprefix\url{https://link.aps.org/doi/10.1103/PhysRevA.72.023613}.

\bibitem[{\citenamefont{Chen and Yan}(2018)}]{Chen:2018}
\bibinfo{author}{\bibfnamefont{T.}~\bibnamefont{Chen}} \bibnamefont{and}
  \bibinfo{author}{\bibfnamefont{B.}~\bibnamefont{Yan}},
  \bibinfo{journal}{Phys. Rev. A} \textbf{\bibinfo{volume}{98}},
  \bibinfo{pages}{063615} (\bibinfo{year}{2018}),
  \urlprefix\url{https://link.aps.org/doi/10.1103/PhysRevA.98.063615}.

\bibitem[{\citenamefont{{Sabari, Subramaniyan} and {Kumar, R.
  Kishor}}(2018)}]{Sabari:2018}
\bibinfo{author}{\bibnamefont{{Sabari, Subramaniyan}}} \bibnamefont{and}
  \bibinfo{author}{\bibnamefont{{Kumar, R. Kishor}}}, \bibinfo{journal}{Eur.
  Phys. J. D} \textbf{\bibinfo{volume}{72}}, \bibinfo{pages}{48}
  (\bibinfo{year}{2018}),
  \urlprefix\url{https://doi.org/10.1140/epjd/e2018-80354-2}.

\bibitem[{\citenamefont{{Cairncross, William} and {Pelster,
  Axel}}(2014)}]{Cairncross:2014}
\bibinfo{author}{\bibnamefont{{Cairncross, William}}} \bibnamefont{and}
  \bibinfo{author}{\bibnamefont{{Pelster, Axel}}}, \bibinfo{journal}{Eur. Phys.
  J. D} \textbf{\bibinfo{volume}{68}}, \bibinfo{pages}{106}
  (\bibinfo{year}{2014}),
  \urlprefix\url{https://doi.org/10.1140/epjd/e2014-40835-x}.

\bibitem[{\citenamefont{Vidanovi\ifmmode~\acute{c}\else \'{c}\fi{}
  et~al.}(2011)\citenamefont{Vidanovi\ifmmode~\acute{c}\else \'{c}\fi{},
  Bala\ifmmode~\check{z}\else \v{z}\fi{}, Al-Jibbouri, and
  Pelster}}]{Vidanovic:2011}
\bibinfo{author}{\bibfnamefont{I.}~\bibnamefont{Vidanovi\ifmmode~\acute{c}\else
  \'{c}\fi{}}},
  \bibinfo{author}{\bibfnamefont{A.}~\bibnamefont{Bala\ifmmode~\check{z}\else
  \v{z}\fi{}}}, \bibinfo{author}{\bibfnamefont{H.}~\bibnamefont{Al-Jibbouri}},
  \bibnamefont{and} \bibinfo{author}{\bibfnamefont{A.}~\bibnamefont{Pelster}},
  \bibinfo{journal}{Phys. Rev. A} \textbf{\bibinfo{volume}{84}},
  \bibinfo{pages}{013618} (\bibinfo{year}{2011}),
  \urlprefix\url{https://link.aps.org/doi/10.1103/PhysRevA.84.013618}.

\bibitem[{\citenamefont{Arnold et~al.}(1989)\citenamefont{Arnold, Weinstein,
  and Vogtmann}}]{Arnold:1989}
\bibinfo{author}{\bibfnamefont{V.~I.} \bibnamefont{Arnold}},
  \bibinfo{author}{\bibfnamefont{A.}~\bibnamefont{Weinstein}},
  \bibnamefont{and} \bibinfo{author}{\bibfnamefont{K.}~\bibnamefont{Vogtmann}},
  \emph{\bibinfo{title}{Mathematical Methods of Classical Mechanics, Graduate
  Texts in Mathematics}}, vol.~\bibinfo{volume}{60}
  (\bibinfo{publisher}{Springer,Berlin}, \bibinfo{year}{1989}).

\bibitem[{\citenamefont{Landau and Lifshitz}(1976)}]{Landau:1976}
\bibinfo{author}{\bibfnamefont{L.~D.} \bibnamefont{Landau}} \bibnamefont{and}
  \bibinfo{author}{\bibfnamefont{E.}~\bibnamefont{Lifshitz}},
  \emph{\bibinfo{title}{Mechanics, Course of Theoretical Physics}},
  vol.~\bibinfo{volume}{1} (\bibinfo{publisher}{Butterworth-Heinemann, Oxford},
  \bibinfo{year}{1976}).

\bibitem[{\citenamefont{Armaroli and Biancalana}(2013)}]{Armaroli:2013}
\bibinfo{author}{\bibfnamefont{A.}~\bibnamefont{Armaroli}} \bibnamefont{and}
  \bibinfo{author}{\bibfnamefont{F.}~\bibnamefont{Biancalana}},
  \bibinfo{journal}{Phys. Rev. A} \textbf{\bibinfo{volume}{87}},
  \bibinfo{pages}{063848} (\bibinfo{year}{2013}),
  \urlprefix\url{https://link.aps.org/doi/10.1103/PhysRevA.87.063848}.

\bibitem[{\citenamefont{Beato and Palacios-Laloy}(2020)}]{Beato:2020}
\bibinfo{author}{\bibfnamefont{F.}~\bibnamefont{Beato}} \bibnamefont{and}
  \bibinfo{author}{\bibfnamefont{A.}~\bibnamefont{Palacios-Laloy}},
  \bibinfo{journal}{EPJ Quantum Technology} \textbf{\bibinfo{volume}{7}},
  \bibinfo{pages}{1} (\bibinfo{year}{2020}).

\bibitem[{\citenamefont{Fomin et~al.}(1997)\citenamefont{Fomin, Shalaev, and
  Shantsev}}]{Fomin:1997}
\bibinfo{author}{\bibfnamefont{N.~V.} \bibnamefont{Fomin}},
  \bibinfo{author}{\bibfnamefont{O.~L.} \bibnamefont{Shalaev}},
  \bibnamefont{and} \bibinfo{author}{\bibfnamefont{D.~V.}
  \bibnamefont{Shantsev}}, \bibinfo{journal}{Journal of Applied Physics}
  \textbf{\bibinfo{volume}{81}}, \bibinfo{pages}{8091} (\bibinfo{year}{1997}),
  \eprint{https://doi.org/10.1063/1.365417},
  \urlprefix\url{https://doi.org/10.1063/1.365417}.

\bibitem[{\citenamefont{Calvo}(1999)}]{Calvo:1999}
\bibinfo{author}{\bibfnamefont{M.}~\bibnamefont{Calvo}},
  \bibinfo{journal}{Phys. Rev. B} \textbf{\bibinfo{volume}{60}},
  \bibinfo{pages}{10953} (\bibinfo{year}{1999}),
  \urlprefix\url{https://link.aps.org/doi/10.1103/PhysRevB.60.10953}.

\bibitem[{\citenamefont{Hackenbroich et~al.}(1998)\citenamefont{Hackenbroich,
  Rosenow, and Weidenm\"uller}}]{Hackenbroich:1998}
\bibinfo{author}{\bibfnamefont{G.}~\bibnamefont{Hackenbroich}},
  \bibinfo{author}{\bibfnamefont{B.}~\bibnamefont{Rosenow}}, \bibnamefont{and}
  \bibinfo{author}{\bibfnamefont{H.~A.} \bibnamefont{Weidenm\"uller}},
  \bibinfo{journal}{Phys. Rev. Lett.} \textbf{\bibinfo{volume}{81}},
  \bibinfo{pages}{5896} (\bibinfo{year}{1998}),
  \urlprefix\url{https://link.aps.org/doi/10.1103/PhysRevLett.81.5896}.

\bibitem[{\citenamefont{Matera et~al.}(1993)\citenamefont{Matera, Mecozzi,
  Romagnoli, and Settembre}}]{Matera:1993}
\bibinfo{author}{\bibfnamefont{F.}~\bibnamefont{Matera}},
  \bibinfo{author}{\bibfnamefont{A.}~\bibnamefont{Mecozzi}},
  \bibinfo{author}{\bibfnamefont{M.}~\bibnamefont{Romagnoli}},
  \bibnamefont{and}
  \bibinfo{author}{\bibfnamefont{M.}~\bibnamefont{Settembre}},
  \bibinfo{journal}{Opt. Lett.} \textbf{\bibinfo{volume}{18}},
  \bibinfo{pages}{1499} (\bibinfo{year}{1993}),
  \urlprefix\url{http://ol.osa.org/abstract.cfm?URI=ol-18-18-1499}.

\bibitem[{\citenamefont{Bismut et~al.}(2010)\citenamefont{Bismut, Pasquiou,
  Mar\'echal, Pedri, Vernac, Gorceix, and Laburthe-Tolra}}]{Bismut:2010}
\bibinfo{author}{\bibfnamefont{G.}~\bibnamefont{Bismut}},
  \bibinfo{author}{\bibfnamefont{B.}~\bibnamefont{Pasquiou}},
  \bibinfo{author}{\bibfnamefont{E.}~\bibnamefont{Mar\'echal}},
  \bibinfo{author}{\bibfnamefont{P.}~\bibnamefont{Pedri}},
  \bibinfo{author}{\bibfnamefont{L.}~\bibnamefont{Vernac}},
  \bibinfo{author}{\bibfnamefont{O.}~\bibnamefont{Gorceix}}, \bibnamefont{and}
  \bibinfo{author}{\bibfnamefont{B.}~\bibnamefont{Laburthe-Tolra}},
  \bibinfo{journal}{Phys. Rev. Lett.} \textbf{\bibinfo{volume}{105}},
  \bibinfo{pages}{040404} (\bibinfo{year}{2010}),
  \urlprefix\url{https://link.aps.org/doi/10.1103/PhysRevLett.105.040404}.

\bibitem[{\citenamefont{G\'oral and Santos}(2002)}]{Goral:2002}
\bibinfo{author}{\bibfnamefont{K.}~\bibnamefont{G\'oral}} \bibnamefont{and}
  \bibinfo{author}{\bibfnamefont{L.}~\bibnamefont{Santos}},
  \bibinfo{journal}{Phys. Rev. A} \textbf{\bibinfo{volume}{66}},
  \bibinfo{pages}{023613} (\bibinfo{year}{2002}),
  \urlprefix\url{https://link.aps.org/doi/10.1103/PhysRevA.66.023613}.

\bibitem[{\citenamefont{Mishra and Nath}(2016)}]{Mishra:2016}
\bibinfo{author}{\bibfnamefont{C.}~\bibnamefont{Mishra}} \bibnamefont{and}
  \bibinfo{author}{\bibfnamefont{R.}~\bibnamefont{Nath}},
  \bibinfo{journal}{Phys. Rev. A} \textbf{\bibinfo{volume}{94}},
  \bibinfo{pages}{033633} (\bibinfo{year}{2016}),
  \urlprefix\url{https://link.aps.org/doi/10.1103/PhysRevA.94.033633}.

\bibitem[{\citenamefont{W\"achtler and Santos}(2016)}]{Waechtler:2016}
\bibinfo{author}{\bibfnamefont{F.}~\bibnamefont{W\"achtler}} \bibnamefont{and}
  \bibinfo{author}{\bibfnamefont{L.}~\bibnamefont{Santos}},
  \bibinfo{journal}{Phys. Rev. A} \textbf{\bibinfo{volume}{94}},
  \bibinfo{pages}{043618} (\bibinfo{year}{2016}),
  \urlprefix\url{https://link.aps.org/doi/10.1103/PhysRevA.94.043618}.

\bibitem[{\citenamefont{Schulz et~al.}(2015)\citenamefont{Schulz, Sala, and
  Saenz}}]{Schulz:2015}
\bibinfo{author}{\bibfnamefont{B.}~\bibnamefont{Schulz}},
  \bibinfo{author}{\bibfnamefont{S.}~\bibnamefont{Sala}}, \bibnamefont{and}
  \bibinfo{author}{\bibfnamefont{A.}~\bibnamefont{Saenz}},
  \bibinfo{journal}{New Journal of Physics} \textbf{\bibinfo{volume}{17}},
  \bibinfo{pages}{065002} (\bibinfo{year}{2015}),
  \urlprefix\url{https://doi.org/10.1088%2F1367-2630%2F17%2F6%2F065002}.

\bibitem[{\citenamefont{Lima and Pelster}(2011)}]{Lima:2011}
\bibinfo{author}{\bibfnamefont{A.~R.~P.} \bibnamefont{Lima}} \bibnamefont{and}
  \bibinfo{author}{\bibfnamefont{A.}~\bibnamefont{Pelster}},
  \bibinfo{journal}{Phys. Rev. A} \textbf{\bibinfo{volume}{84}},
  \bibinfo{pages}{041604} (\bibinfo{year}{2011}),
  \urlprefix\url{https://link.aps.org/doi/10.1103/PhysRevA.84.041604}.

\bibitem[{\citenamefont{Ancilotto and Toigo}(2014)}]{Ancilotto:2014}
\bibinfo{author}{\bibfnamefont{F.}~\bibnamefont{Ancilotto}} \bibnamefont{and}
  \bibinfo{author}{\bibfnamefont{F.}~\bibnamefont{Toigo}},
  \bibinfo{journal}{Phys. Rev. A} \textbf{\bibinfo{volume}{89}},
  \bibinfo{pages}{023617} (\bibinfo{year}{2014}),
  \urlprefix\url{https://link.aps.org/doi/10.1103/PhysRevA.89.023617}.

\bibitem[{\citenamefont{Wilson and Bohn}(2011)}]{Wilson:2011}
\bibinfo{author}{\bibfnamefont{R.~M.} \bibnamefont{Wilson}} \bibnamefont{and}
  \bibinfo{author}{\bibfnamefont{J.~L.} \bibnamefont{Bohn}},
  \bibinfo{journal}{Phys. Rev. A} \textbf{\bibinfo{volume}{83}},
  \bibinfo{pages}{023623} (\bibinfo{year}{2011}),
  \urlprefix\url{https://link.aps.org/doi/10.1103/PhysRevA.83.023623}.

\bibitem[{\citenamefont{Agrawal}(1987)}]{Agrawal:1987}
\bibinfo{author}{\bibfnamefont{G.~P.} \bibnamefont{Agrawal}},
  \bibinfo{journal}{Phys. Rev. Lett.} \textbf{\bibinfo{volume}{59}},
  \bibinfo{pages}{880} (\bibinfo{year}{1987}),
  \urlprefix\url{https://link.aps.org/doi/10.1103/PhysRevLett.59.880}.

\bibitem[{\citenamefont{Abdullaev et~al.}(1996)\citenamefont{Abdullaev,
  Darmanyan, Kobyakov, and Lederer}}]{Abdullaev:1996}
\bibinfo{author}{\bibfnamefont{F.}~\bibnamefont{Abdullaev}},
  \bibinfo{author}{\bibfnamefont{S.}~\bibnamefont{Darmanyan}},
  \bibinfo{author}{\bibfnamefont{A.}~\bibnamefont{Kobyakov}}, \bibnamefont{and}
  \bibinfo{author}{\bibfnamefont{F.}~\bibnamefont{Lederer}},
  \bibinfo{journal}{Physics Letters A} \textbf{\bibinfo{volume}{220}},
  \bibinfo{pages}{213 } (\bibinfo{year}{1996}), ISSN \bibinfo{issn}{0375-9601},
  \urlprefix\url{http://www.sciencedirect.com/science/article/pii/037596019600504X}.

\bibitem[{\citenamefont{Ambomo et~al.}(2008)\citenamefont{Ambomo, Ngabireng,
  Dinda, Labruy\`{e}re, Porsezian, and Kalithasan}}]{Ambomo:2008}
\bibinfo{author}{\bibfnamefont{S.}~\bibnamefont{Ambomo}},
  \bibinfo{author}{\bibfnamefont{C.~M.} \bibnamefont{Ngabireng}},
  \bibinfo{author}{\bibfnamefont{P.~T.} \bibnamefont{Dinda}},
  \bibinfo{author}{\bibfnamefont{A.}~\bibnamefont{Labruy\`{e}re}},
  \bibinfo{author}{\bibfnamefont{K.}~\bibnamefont{Porsezian}},
  \bibnamefont{and}
  \bibinfo{author}{\bibfnamefont{B.}~\bibnamefont{Kalithasan}},
  \bibinfo{journal}{J. Opt. Soc. Am. B} \textbf{\bibinfo{volume}{25}},
  \bibinfo{pages}{425} (\bibinfo{year}{2008}),
  \urlprefix\url{http://josab.osa.org/abstract.cfm?URI=josab-25-3-425}.

\bibitem[{\citenamefont{{Michael C. Garrett, Adrian Ratnapala, Eikbert D. van
  Ooijen, Christopher J. Vale, Kristian Weegink, Sebastian K. Schnelle, Otto
  Vainio, Norman R. Heckenberg, Halina Rubinsztein-Dunlop, and Matthew J.
  Davis}}(2011)}]{Garrett:2011}
\bibinfo{author}{\bibnamefont{{Michael C. Garrett, Adrian Ratnapala, Eikbert D.
  van Ooijen, Christopher J. Vale, Kristian Weegink, Sebastian K. Schnelle,
  Otto Vainio, Norman R. Heckenberg, Halina Rubinsztein-Dunlop, and Matthew J.
  Davis}}}, \bibinfo{journal}{Phys. Rev. A} \textbf{\bibinfo{volume}{83}},
  \bibinfo{pages}{013630} (\bibinfo{year}{2011}),
  \urlprefix\url{https://doi.org/10.1103/PhysRevA.83.013630}.

\bibitem[{\citenamefont{Clark et~al.}(2015)\citenamefont{Clark, Ha, Xu, and
  Chin}}]{Clark:2015}
\bibinfo{author}{\bibfnamefont{L.~W.} \bibnamefont{Clark}},
  \bibinfo{author}{\bibfnamefont{L.-C.} \bibnamefont{Ha}},
  \bibinfo{author}{\bibfnamefont{C.-Y.} \bibnamefont{Xu}}, \bibnamefont{and}
  \bibinfo{author}{\bibfnamefont{C.}~\bibnamefont{Chin}},
  \bibinfo{journal}{Phys. Rev. Lett.} \textbf{\bibinfo{volume}{115}},
  \bibinfo{pages}{155301} (\bibinfo{year}{2015}).

\bibitem[{\citenamefont{Oppenheim and Verghese}(2015)}]{Oppenheim:2015}
\bibinfo{author}{\bibfnamefont{A.~V.} \bibnamefont{Oppenheim}}
  \bibnamefont{and} \bibinfo{author}{\bibfnamefont{G.~C.}
  \bibnamefont{Verghese}}, \emph{\bibinfo{title}{Signals, Systems, and
  Inference}} (\bibinfo{publisher}{Prentice Hall Signal Processing Series,
  Pearson}, \bibinfo{year}{2015}), \bibinfo{edition}{1st} ed.

\bibitem[{\citenamefont{Muruganandam and Adhikari}(2012)}]{Muruganandam:2012}
\bibinfo{author}{\bibfnamefont{P.}~\bibnamefont{Muruganandam}}
  \bibnamefont{and} \bibinfo{author}{\bibfnamefont{S.~K.}
  \bibnamefont{Adhikari}}, \bibinfo{journal}{Las. Phys.}
  \textbf{\bibinfo{volume}{22}}, \bibinfo{pages}{813} (\bibinfo{year}{2012}).

\bibitem[{\citenamefont{{Roger Sakhel and Asaad Sakhel}}(2017)}]{Sakhel:2017}
\bibinfo{author}{\bibnamefont{{Roger Sakhel and Asaad Sakhel}}},
  \bibinfo{journal}{J. Phys. B: At. Mol. Opt. Phys.}
  \textbf{\bibinfo{volume}{50}}, \bibinfo{pages}{105301}
  (\bibinfo{year}{2017}).

\bibitem[{\citenamefont{{V. M. Perez-Garcia, H. Michinel, J. I. Cirac, M.
  Lewenstein, and P. Z{\"o}ller}}(1996)}]{Perez-Garcia:1996}
\bibinfo{author}{\bibnamefont{{V. M. Perez-Garcia, H. Michinel, J. I. Cirac, M.
  Lewenstein, and P. Z{\"o}ller}}}, \bibinfo{journal}{Phys. Rev. Lett.}
  \textbf{\bibinfo{volume}{77}}, \bibinfo{pages}{5320} (\bibinfo{year}{1996}).

\bibitem[{\citenamefont{{V. M. Perez-Garcia, H. Michinel, J. I. Cirac, M.
  Lewenstein, and P. Z{\"o}ller}}(1997)}]{Perez-Garcia:1997}
\bibinfo{author}{\bibnamefont{{V. M. Perez-Garcia, H. Michinel, J. I. Cirac, M.
  Lewenstein, and P. Z{\"o}ller}}}, \bibinfo{journal}{Phys. Rev. A}
  \textbf{\bibinfo{volume}{56}}, \bibinfo{pages}{1424} (\bibinfo{year}{1997}).

\bibitem[{\citenamefont{{Hamid Al-Jibbouri, Ivana Vidanovic, Antun Balaz, and
  Axel Pelster}}(2013)}]{Al-Jibbouri:2013}
\bibinfo{author}{\bibnamefont{{Hamid Al-Jibbouri, Ivana Vidanovic, Antun Balaz,
  and Axel Pelster}}}, \bibinfo{journal}{J. Phys. B: At. Mol. Opt. Phys.}
  \textbf{\bibinfo{volume}{46}}, \bibinfo{pages}{065303}
  (\bibinfo{year}{2013}).

\bibitem[{\citenamefont{{G. M. Falco}}(2009)}]{Falco:2009}
\bibinfo{author}{\bibnamefont{{G. M. Falco}}}, \bibinfo{journal}{J. Phys. B:
  At. Mol. Opt. Phys.} \textbf{\bibinfo{volume}{42}}, \bibinfo{pages}{215303}
  (\bibinfo{year}{2009}).

\bibitem[{\citenamefont{{Qi Wei, Liang Zhao-Xin, and Zhang
  Zhi-Dong}}(2013)}]{Wei:2013}
\bibinfo{author}{\bibnamefont{{Qi Wei, Liang Zhao-Xin, and Zhang Zhi-Dong}}},
  \bibinfo{journal}{Chinese Physics Letters} \textbf{\bibinfo{volume}{30}},
  \bibinfo{pages}{060303} (\bibinfo{year}{2013}).

\bibitem[{\citenamefont{{R. Kishor Kumar, Luis E. Young-S., Du{\v{s}}san
  Vudragovi{\'{c}}, Antun Bala{\v{z}}, Paulsamy Muruganandam,
  S.K.Adhikari}}(2015)}]{Kumar:2015}
\bibinfo{author}{\bibnamefont{{R. Kishor Kumar, Luis E. Young-S., Du{\v{s}}san
  Vudragovi{\'{c}}, Antun Bala{\v{z}}, Paulsamy Muruganandam, S.K.Adhikari}}},
  \bibinfo{journal}{Computer Physics Communications}
  \textbf{\bibinfo{volume}{195}}, \bibinfo{pages}{117} (\bibinfo{year}{2015}),
  \urlprefix\url{https://www.sciencedirect.com/science/article/pii/S0010465515001344}.

\bibitem[{\citenamefont{Bruce et~al.}(2011)\citenamefont{Bruce, Bromley,
  Smirne, Torralbo-Campo, and Cassettari}}]{Bruce:2011b}
\bibinfo{author}{\bibfnamefont{G.~D.} \bibnamefont{Bruce}},
  \bibinfo{author}{\bibfnamefont{S.~L.} \bibnamefont{Bromley}},
  \bibinfo{author}{\bibfnamefont{G.}~\bibnamefont{Smirne}},
  \bibinfo{author}{\bibfnamefont{L.}~\bibnamefont{Torralbo-Campo}},
  \bibnamefont{and}
  \bibinfo{author}{\bibfnamefont{D.}~\bibnamefont{Cassettari}},
  \bibinfo{journal}{Phys. Rev. A} \textbf{\bibinfo{volume}{84}},
  \bibinfo{pages}{053410} (\bibinfo{year}{2011}),
  \urlprefix\url{https://link.aps.org/doi/10.1103/PhysRevA.84.053410}.

\bibitem[{\citenamefont{Hammes et~al.}(2002)\citenamefont{Hammes, Rychtarik,
  N\"agerl, and Grimm}}]{Hammes:2002}
\bibinfo{author}{\bibfnamefont{M.}~\bibnamefont{Hammes}},
  \bibinfo{author}{\bibfnamefont{D.}~\bibnamefont{Rychtarik}},
  \bibinfo{author}{\bibfnamefont{H.-C.} \bibnamefont{N\"agerl}},
  \bibnamefont{and} \bibinfo{author}{\bibfnamefont{R.}~\bibnamefont{Grimm}},
  \bibinfo{journal}{Phys. Rev. A} \textbf{\bibinfo{volume}{66}},
  \bibinfo{pages}{051401} (\bibinfo{year}{2002}),
  \urlprefix\url{https://link.aps.org/doi/10.1103/PhysRevA.66.051401}.

\bibitem[{\citenamefont{Tuchendler et~al.}(2008)\citenamefont{Tuchendler,
  Lance, Browaeys, Sortais, and Grangier}}]{Tuchendler:2008}
\bibinfo{author}{\bibfnamefont{C.}~\bibnamefont{Tuchendler}},
  \bibinfo{author}{\bibfnamefont{A.~M.} \bibnamefont{Lance}},
  \bibinfo{author}{\bibfnamefont{A.}~\bibnamefont{Browaeys}},
  \bibinfo{author}{\bibfnamefont{Y.~R.~P.} \bibnamefont{Sortais}},
  \bibnamefont{and} \bibinfo{author}{\bibfnamefont{P.}~\bibnamefont{Grangier}},
  \bibinfo{journal}{Phys. Rev. A} \textbf{\bibinfo{volume}{78}},
  \bibinfo{pages}{033425} (\bibinfo{year}{2008}),
  \urlprefix\url{https://link.aps.org/doi/10.1103/PhysRevA.78.033425}.

\bibitem[{\citenamefont{Stamper-Kurn et~al.}(1998)\citenamefont{Stamper-Kurn,
  Miesner, Chikkatur, Inouye, Stenger, and Ketterle}}]{Stamper-Kurn:1998}
\bibinfo{author}{\bibfnamefont{D.~M.} \bibnamefont{Stamper-Kurn}},
  \bibinfo{author}{\bibfnamefont{H.-J.} \bibnamefont{Miesner}},
  \bibinfo{author}{\bibfnamefont{A.~P.} \bibnamefont{Chikkatur}},
  \bibinfo{author}{\bibfnamefont{S.}~\bibnamefont{Inouye}},
  \bibinfo{author}{\bibfnamefont{J.}~\bibnamefont{Stenger}}, \bibnamefont{and}
  \bibinfo{author}{\bibfnamefont{W.}~\bibnamefont{Ketterle}},
  \bibinfo{journal}{Phys. Rev. Lett.} \textbf{\bibinfo{volume}{81}},
  \bibinfo{pages}{2194} (\bibinfo{year}{1998}),
  \urlprefix\url{https://link.aps.org/doi/10.1103/PhysRevLett.81.2194}.

\bibitem[{\citenamefont{Comparat et~al.}(2006)\citenamefont{Comparat, Fioretti,
  Stern, Dimova, Tolra, and Pillet}}]{Comparat:2006}
\bibinfo{author}{\bibfnamefont{D.}~\bibnamefont{Comparat}},
  \bibinfo{author}{\bibfnamefont{A.}~\bibnamefont{Fioretti}},
  \bibinfo{author}{\bibfnamefont{G.}~\bibnamefont{Stern}},
  \bibinfo{author}{\bibfnamefont{E.}~\bibnamefont{Dimova}},
  \bibinfo{author}{\bibfnamefont{B.~L.} \bibnamefont{Tolra}}, \bibnamefont{and}
  \bibinfo{author}{\bibfnamefont{P.}~\bibnamefont{Pillet}},
  \bibinfo{journal}{Phys. Rev. A} \textbf{\bibinfo{volume}{73}},
  \bibinfo{pages}{043410} (\bibinfo{year}{2006}),
  \urlprefix\url{https://link.aps.org/doi/10.1103/PhysRevA.73.043410}.

\bibitem[{\citenamefont{Jacob et~al.}(2011)\citenamefont{Jacob, Mimoun, Sarlo,
  Weitz, Dalibard, and Gerbier}}]{Jacob:2011}
\bibinfo{author}{\bibfnamefont{D.}~\bibnamefont{Jacob}},
  \bibinfo{author}{\bibfnamefont{E.}~\bibnamefont{Mimoun}},
  \bibinfo{author}{\bibfnamefont{L.~D.} \bibnamefont{Sarlo}},
  \bibinfo{author}{\bibfnamefont{M.}~\bibnamefont{Weitz}},
  \bibinfo{author}{\bibfnamefont{J.}~\bibnamefont{Dalibard}}, \bibnamefont{and}
  \bibinfo{author}{\bibfnamefont{F.}~\bibnamefont{Gerbier}},
  \bibinfo{journal}{New J. Phys.} \textbf{\bibinfo{volume}{13}},
  \bibinfo{pages}{065022} (\bibinfo{year}{2011}),
  \urlprefix\url{https://doi.org/10.1088%2F1367-2630%2F13%2F6%2F065022}.

\bibitem[{\citenamefont{Gustavson et~al.}(2001)\citenamefont{Gustavson,
  Chikkatur, Leanhardt, G\"orlitz, Gupta, Pritchard, and
  Ketterle}}]{Gustavson:2001}
\bibinfo{author}{\bibfnamefont{T.~L.} \bibnamefont{Gustavson}},
  \bibinfo{author}{\bibfnamefont{A.~P.} \bibnamefont{Chikkatur}},
  \bibinfo{author}{\bibfnamefont{A.~E.} \bibnamefont{Leanhardt}},
  \bibinfo{author}{\bibfnamefont{A.}~\bibnamefont{G\"orlitz}},
  \bibinfo{author}{\bibfnamefont{S.}~\bibnamefont{Gupta}},
  \bibinfo{author}{\bibfnamefont{D.~E.} \bibnamefont{Pritchard}},
  \bibnamefont{and} \bibinfo{author}{\bibfnamefont{W.}~\bibnamefont{Ketterle}},
  \bibinfo{journal}{Phys. Rev. Lett.} \textbf{\bibinfo{volume}{88}},
  \bibinfo{pages}{020401} (\bibinfo{year}{2001}),
  \urlprefix\url{https://link.aps.org/doi/10.1103/PhysRevLett.88.020401}.

\bibitem[{\citenamefont{Barrett et~al.}(2001)\citenamefont{Barrett, Sauer, and
  Chapman}}]{Barrett:2001}
\bibinfo{author}{\bibfnamefont{M.~D.} \bibnamefont{Barrett}},
  \bibinfo{author}{\bibfnamefont{J.~A.} \bibnamefont{Sauer}}, \bibnamefont{and}
  \bibinfo{author}{\bibfnamefont{M.~S.} \bibnamefont{Chapman}},
  \bibinfo{journal}{Phys. Rev. Lett.} \textbf{\bibinfo{volume}{87}},
  \bibinfo{pages}{010404} (\bibinfo{year}{2001}),
  \urlprefix\url{https://link.aps.org/doi/10.1103/PhysRevLett.87.010404}.

\bibitem[{\citenamefont{Schulz et~al.}(2007)\citenamefont{Schulz, Crepaz,
  Schmidt-Kaler, Eschner, and Blatt}}]{Schulz:2007}
\bibinfo{author}{\bibfnamefont{M.}~\bibnamefont{Schulz}},
  \bibinfo{author}{\bibfnamefont{H.}~\bibnamefont{Crepaz}},
  \bibinfo{author}{\bibfnamefont{F.}~\bibnamefont{Schmidt-Kaler}},
  \bibinfo{author}{\bibfnamefont{J.}~\bibnamefont{Eschner}}, \bibnamefont{and}
  \bibinfo{author}{\bibfnamefont{R.}~\bibnamefont{Blatt}}, \bibinfo{journal}{J.
  Mod. Opt.} \textbf{\bibinfo{volume}{54}}, \bibinfo{pages}{1619}
  (\bibinfo{year}{2007}), \eprint{https://doi.org/10.1080/09500340600861740},
  \urlprefix\url{https://doi.org/10.1080/09500340600861740}.

\bibitem[{\citenamefont{Proukakis et~al.}(2006)\citenamefont{Proukakis,
  Schmiedmayer, and Stoof}}]{Proukakis:2006}
\bibinfo{author}{\bibfnamefont{N.~P.} \bibnamefont{Proukakis}},
  \bibinfo{author}{\bibfnamefont{J.}~\bibnamefont{Schmiedmayer}},
  \bibnamefont{and} \bibinfo{author}{\bibfnamefont{H.~T.~C.}
  \bibnamefont{Stoof}}, \bibinfo{journal}{Phys. Rev. A}
  \textbf{\bibinfo{volume}{73}}, \bibinfo{pages}{053603}
  (\bibinfo{year}{2006}),
  \urlprefix\url{https://link.aps.org/doi/10.1103/PhysRevA.73.053603}.

\bibitem[{\citenamefont{Diener et~al.}(2002)\citenamefont{Diener, {Biao Wu},
  Raizen, and Niu}}]{Diener:2002}
\bibinfo{author}{\bibfnamefont{R.~B.} \bibnamefont{Diener}},
  \bibinfo{author}{\bibnamefont{{Biao Wu}}},
  \bibinfo{author}{\bibfnamefont{M.~G.} \bibnamefont{Raizen}},
  \bibnamefont{and} \bibinfo{author}{\bibfnamefont{Q.}~\bibnamefont{Niu}},
  \bibinfo{journal}{Phys. Rev. Lett.} \textbf{\bibinfo{volume}{89}},
  \bibinfo{pages}{070401} (\bibinfo{year}{2002}),
  \urlprefix\url{https://link.aps.org/doi/10.1103/PhysRevLett.89.070401}.

\bibitem[{\citenamefont{Aioi et~al.}(2011)\citenamefont{Aioi, Kadokura,
  Kishimoto, and Saito}}]{Aioi:2011}
\bibinfo{author}{\bibfnamefont{T.}~\bibnamefont{Aioi}},
  \bibinfo{author}{\bibfnamefont{T.}~\bibnamefont{Kadokura}},
  \bibinfo{author}{\bibfnamefont{T.}~\bibnamefont{Kishimoto}},
  \bibnamefont{and} \bibinfo{author}{\bibfnamefont{H.}~\bibnamefont{Saito}},
  \bibinfo{journal}{Phys. Rev. X} \textbf{\bibinfo{volume}{1}},
  \bibinfo{pages}{021003} (\bibinfo{year}{2011}),
  \urlprefix\url{https://link.aps.org/doi/10.1103/PhysRevX.1.021003}.

\bibitem[{\citenamefont{Uncu et~al.}(2008)\citenamefont{Uncu, Tarhan, Demiralp,
  and M\"ustecaplioglu}}]{Uncu:2008}
\bibinfo{author}{\bibfnamefont{H.}~\bibnamefont{Uncu}},
  \bibinfo{author}{\bibfnamefont{D.}~\bibnamefont{Tarhan}},
  \bibinfo{author}{\bibfnamefont{E.}~\bibnamefont{Demiralp}}, \bibnamefont{and}
  \bibinfo{author}{\bibfnamefont{O.~E.} \bibnamefont{M\"ustecaplioglu}},
  \bibinfo{journal}{Las. Phys.} \textbf{\bibinfo{volume}{18}},
  \bibinfo{pages}{331} (\bibinfo{year}{2008}),
  \urlprefix\url{https://doi.org/10.1134/S1054660X08030237}.

\bibitem[{\citenamefont{Weitenberg et~al.}(2011)\citenamefont{Weitenberg, Kuhr,
  M\o{}lmer, and Sherson}}]{Weitenberg:2011}
\bibinfo{author}{\bibfnamefont{C.}~\bibnamefont{Weitenberg}},
  \bibinfo{author}{\bibfnamefont{S.}~\bibnamefont{Kuhr}},
  \bibinfo{author}{\bibfnamefont{K.}~\bibnamefont{M\o{}lmer}},
  \bibnamefont{and} \bibinfo{author}{\bibfnamefont{J.~F.}
  \bibnamefont{Sherson}}, \bibinfo{journal}{Phys. Rev. A}
  \textbf{\bibinfo{volume}{84}}, \bibinfo{pages}{032322}
  (\bibinfo{year}{2011}),
  \urlprefix\url{https://link.aps.org/doi/10.1103/PhysRevA.84.032322}.

\bibitem[{\citenamefont{Carpentier et~al.}(2008)\citenamefont{Carpentier,
  Belmonte-Beitia, Michinel, and Rodas-Verde}}]{Carpentier:2008}
\bibinfo{author}{\bibfnamefont{A.~V.} \bibnamefont{Carpentier}},
  \bibinfo{author}{\bibfnamefont{J.}~\bibnamefont{Belmonte-Beitia}},
  \bibinfo{author}{\bibfnamefont{H.}~\bibnamefont{Michinel}}, \bibnamefont{and}
  \bibinfo{author}{\bibfnamefont{M.~I.} \bibnamefont{Rodas-Verde}},
  \bibinfo{journal}{J. of Mod. Opt.} \textbf{\bibinfo{volume}{55}},
  \bibinfo{pages}{2819} (\bibinfo{year}{2008}),
  \eprint{https://doi.org/10.1080/09500340802209763},
  \urlprefix\url{https://doi.org/10.1080/09500340802209763}.

\bibitem[{\citenamefont{Parker et~al.}(2003)\citenamefont{Parker, Proukakis,
  Leadbeater, and Adams}}]{Parker:2003}
\bibinfo{author}{\bibfnamefont{N.~G.} \bibnamefont{Parker}},
  \bibinfo{author}{\bibfnamefont{N.~P.} \bibnamefont{Proukakis}},
  \bibinfo{author}{\bibfnamefont{M.}~\bibnamefont{Leadbeater}},
  \bibnamefont{and} \bibinfo{author}{\bibfnamefont{C.~S.} \bibnamefont{Adams}},
  \bibinfo{journal}{Phys. Rev. Lett.} \textbf{\bibinfo{volume}{90}},
  \bibinfo{pages}{220401} (\bibinfo{year}{2003}),
  \urlprefix\url{https://link.aps.org/doi/10.1103/PhysRevLett.90.220401}.

\bibitem[{\citenamefont{Pethick and H.}(2008)}]{Pethick:2008}
\bibinfo{author}{\bibfnamefont{C.~J.} \bibnamefont{Pethick}} \bibnamefont{and}
  \bibinfo{author}{\bibfnamefont{S.}~\bibnamefont{H.}},
  \emph{\bibinfo{title}{Bose-Einstein Condensation in Dilute Gases}}
  (\bibinfo{publisher}{Cambridge University Press, Cambridge},
  \bibinfo{year}{2008}), \bibinfo{edition}{2nd} ed.

\bibitem[{\citenamefont{Lahaye et~al.}(2009)\citenamefont{Lahaye, Menotti,
  Santos, Lewenstein, and Pfau}}]{Lahaye:2009}
\bibinfo{author}{\bibfnamefont{T.}~\bibnamefont{Lahaye}},
  \bibinfo{author}{\bibfnamefont{C.}~\bibnamefont{Menotti}},
  \bibinfo{author}{\bibfnamefont{L.}~\bibnamefont{Santos}},
  \bibinfo{author}{\bibfnamefont{M.}~\bibnamefont{Lewenstein}},
  \bibnamefont{and} \bibinfo{author}{\bibfnamefont{T.}~\bibnamefont{Pfau}},
  \bibinfo{journal}{Reports on Progress in Physics}
  \textbf{\bibinfo{volume}{72}}, \bibinfo{pages}{126401}
  (\bibinfo{year}{2009}),
  \urlprefix\url{https://doi.org/10.1088%2F0034-4885%2F72%2F12%2F126401}.

\bibitem[{\citenamefont{O'Dell et~al.}(2004)\citenamefont{O'Dell, Giovanazzi,
  and Eberlein}}]{Duncan:2004}
\bibinfo{author}{\bibfnamefont{D.~H.~J.} \bibnamefont{O'Dell}},
  \bibinfo{author}{\bibfnamefont{S.}~\bibnamefont{Giovanazzi}},
  \bibnamefont{and} \bibinfo{author}{\bibfnamefont{C.}~\bibnamefont{Eberlein}},
  \bibinfo{journal}{Phys. Rev. Lett.} \textbf{\bibinfo{volume}{92}},
  \bibinfo{pages}{250401} (\bibinfo{year}{2004}),
  \urlprefix\url{https://link.aps.org/doi/10.1103/PhysRevLett.92.250401}.

\bibitem[{\citenamefont{Yi and You}(2000)}]{Yi:2000}
\bibinfo{author}{\bibfnamefont{S.}~\bibnamefont{Yi}} \bibnamefont{and}
  \bibinfo{author}{\bibfnamefont{L.}~\bibnamefont{You}},
  \bibinfo{journal}{Phys. Rev. A} \textbf{\bibinfo{volume}{61}},
  \bibinfo{pages}{041604} (\bibinfo{year}{2000}),
  \urlprefix\url{https://link.aps.org/doi/10.1103/PhysRevA.61.041604}.

\bibitem[{\citenamefont{Bohn et~al.}(2009)\citenamefont{Bohn, Cavagnero, and
  Ticknor}}]{Bohn:2009}
\bibinfo{author}{\bibfnamefont{J.~L.} \bibnamefont{Bohn}},
  \bibinfo{author}{\bibfnamefont{M.}~\bibnamefont{Cavagnero}},
  \bibnamefont{and} \bibinfo{author}{\bibfnamefont{C.}~\bibnamefont{Ticknor}},
  \bibinfo{journal}{New Journal of Physics} \textbf{\bibinfo{volume}{11}},
  \bibinfo{pages}{055039} (\bibinfo{year}{2009}),
  \urlprefix\url{https://doi.org/10.1088%2F1367-2630%2F11%2F5%2F055039}.

\bibitem[{\citenamefont{G\'oral et~al.}(2000)\citenamefont{G\'oral, Rza\ifmmode
  \mbox{\c{}}\else \c{}\fi{}\ifmmode~\dot{z}\else \.{z}\fi{}ewski, and
  Pfau}}]{Goral:2000}
\bibinfo{author}{\bibfnamefont{K.}~\bibnamefont{G\'oral}},
  \bibinfo{author}{\bibfnamefont{K.}~\bibnamefont{Rza\ifmmode \mbox{\c{}}\else
  \c{}\fi{}\ifmmode~\dot{z}\else \.{z}\fi{}ewski}}, \bibnamefont{and}
  \bibinfo{author}{\bibfnamefont{T.}~\bibnamefont{Pfau}},
  \bibinfo{journal}{Phys. Rev. A} \textbf{\bibinfo{volume}{61}},
  \bibinfo{pages}{051601} (\bibinfo{year}{2000}),
  \urlprefix\url{https://link.aps.org/doi/10.1103/PhysRevA.61.051601}.

\bibitem[{\citenamefont{Olson et~al.}(2013)\citenamefont{Olson, Whitenack, and
  Chen}}]{Olson:2013}
\bibinfo{author}{\bibfnamefont{A.~J.} \bibnamefont{Olson}},
  \bibinfo{author}{\bibfnamefont{D.~L.} \bibnamefont{Whitenack}},
  \bibnamefont{and} \bibinfo{author}{\bibfnamefont{Y.~P.} \bibnamefont{Chen}},
  \bibinfo{journal}{Phys. Rev. A} \textbf{\bibinfo{volume}{88}},
  \bibinfo{pages}{043609} (\bibinfo{year}{2013}),
  \urlprefix\url{https://link.aps.org/doi/10.1103/PhysRevA.88.043609}.

\bibitem[{\citenamefont{Koch et~al.}(2008)\citenamefont{Koch, Lahaye, Metz,
  Fröhlich, Griesmaier, and Pfau}}]{Koch:2008}
\bibinfo{author}{\bibfnamefont{T.}~\bibnamefont{Koch}},
  \bibinfo{author}{\bibfnamefont{T.}~\bibnamefont{Lahaye}},
  \bibinfo{author}{\bibfnamefont{J.}~\bibnamefont{Metz}},
  \bibinfo{author}{\bibfnamefont{B.}~\bibnamefont{Fröhlich}},
  \bibinfo{author}{\bibfnamefont{A.}~\bibnamefont{Griesmaier}},
  \bibnamefont{and} \bibinfo{author}{\bibfnamefont{T.}~\bibnamefont{Pfau}},
  \bibinfo{journal}{Nat. Phys.} \textbf{\bibinfo{volume}{4}},
  \bibinfo{pages}{218} (\bibinfo{year}{2008}),
  \urlprefix\url{https://doi.org/10.1038/nphys887}.

\bibitem[{\citenamefont{Lu et~al.}(2011)\citenamefont{Lu, Burdick, Youn, and
  Lev}}]{Lu:2011}
\bibinfo{author}{\bibfnamefont{M.}~\bibnamefont{Lu}},
  \bibinfo{author}{\bibfnamefont{N.~Q.} \bibnamefont{Burdick}},
  \bibinfo{author}{\bibfnamefont{S.~H.} \bibnamefont{Youn}}, \bibnamefont{and}
  \bibinfo{author}{\bibfnamefont{B.~L.} \bibnamefont{Lev}},
  \bibinfo{journal}{Phys. Rev. Lett.} \textbf{\bibinfo{volume}{107}},
  \bibinfo{pages}{190401} (\bibinfo{year}{2011}),
  \urlprefix\url{https://link.aps.org/doi/10.1103/PhysRevLett.107.190401}.

\bibitem[{\citenamefont{Youn et~al.}(2010)\citenamefont{Youn, Lu, Ray, and
  Lev}}]{Youn:2010}
\bibinfo{author}{\bibfnamefont{S.~H.} \bibnamefont{Youn}},
  \bibinfo{author}{\bibfnamefont{M.}~\bibnamefont{Lu}},
  \bibinfo{author}{\bibfnamefont{U.}~\bibnamefont{Ray}}, \bibnamefont{and}
  \bibinfo{author}{\bibfnamefont{B.~L.} \bibnamefont{Lev}},
  \bibinfo{journal}{Phys. Rev. A} \textbf{\bibinfo{volume}{82}},
  \bibinfo{pages}{043425} (\bibinfo{year}{2010}),
  \urlprefix\url{https://link.aps.org/doi/10.1103/PhysRevA.82.043425}.

\bibitem[{\citenamefont{Filinov and Bonitz}(2012)}]{Filinov:2012}
\bibinfo{author}{\bibfnamefont{A.}~\bibnamefont{Filinov}} \bibnamefont{and}
  \bibinfo{author}{\bibfnamefont{M.}~\bibnamefont{Bonitz}},
  \bibinfo{journal}{Phys. Rev. A} \textbf{\bibinfo{volume}{86}},
  \bibinfo{pages}{043628} (\bibinfo{year}{2012}),
  \urlprefix\url{https://link.aps.org/doi/10.1103/PhysRevA.86.043628}.

\bibitem[{\citenamefont{Muruganandam and Adhikari}(2009)}]{Muruganandam:2009}
\bibinfo{author}{\bibfnamefont{P.}~\bibnamefont{Muruganandam}}
  \bibnamefont{and} \bibinfo{author}{\bibfnamefont{S.~K.}
  \bibnamefont{Adhikari}}, \bibinfo{journal}{Computer Physics Communications}
  \textbf{\bibinfo{volume}{180}}, \bibinfo{pages}{1888} (\bibinfo{year}{2009}),
  \urlprefix\url{https://www.sciencedirect.com/science/article/pii/S001046550900126X}.

\bibitem[{\citenamefont{{D. Vudragovi\'c, I. Vidanovi\'c, A. Bala\v z, P.
  Muruganandam, and S. K. Adhikari}}(2012)}]{Vudragovic:2012}
\bibinfo{author}{\bibnamefont{{D. Vudragovi\'c, I. Vidanovi\'c, A. Bala\v z, P.
  Muruganandam, and S. K. Adhikari}}}, \bibinfo{journal}{Computer Physics
  Communications} \textbf{\bibinfo{volume}{183}}, \bibinfo{pages}{2021}
  (\bibinfo{year}{2012}),
  \urlprefix\url{https://www.sciencedirect.com/science/article/pii/S0010465512001270}.

\bibitem[{\citenamefont{{V. Lon\v{c}ar, A. Bala\v{z}, A. Bogojevi\ifmmode
  \acute{c}\else \'c{}\fi{}, S. \ifmmode \check{S}\else \v{S}\fi{}krbi\ifmmode
  \acute{c}\else \'c{}\fi{}, P. Muruganandam, and S. K.
  Adhikari}}(2016)}]{Loncar:2016}
\bibinfo{author}{\bibnamefont{{V. Lon\v{c}ar, A. Bala\v{z}, A. Bogojevi\ifmmode
  \acute{c}\else \'c{}\fi{}, S. \ifmmode \check{S}\else \v{S}\fi{}krbi\ifmmode
  \acute{c}\else \'c{}\fi{}, P. Muruganandam, and S. K. Adhikari}}},
  \bibinfo{journal}{Computer Physics Communications}
  \textbf{\bibinfo{volume}{200}}, \bibinfo{pages}{406} (\bibinfo{year}{2016}),
  \urlprefix\url{https://www.sciencedirect.com/science/article/pii/S0010465515004361}.

\bibitem[{\citenamefont{Lon\v{c}ar et~al.}(2016)\citenamefont{Lon\v{c}ar,
  Young-S., \v{S}krbi\'c, Muruganandam, Adhikari, and
  Bala\v{z}}}]{Loncar:2016b}
\bibinfo{author}{\bibfnamefont{V.}~\bibnamefont{Lon\v{c}ar}},
  \bibinfo{author}{\bibfnamefont{L.~E.} \bibnamefont{Young-S.}},
  \bibinfo{author}{\bibfnamefont{S.}~\bibnamefont{\v{S}krbi\'c}},
  \bibinfo{author}{\bibfnamefont{P.}~\bibnamefont{Muruganandam}},
  \bibinfo{author}{\bibfnamefont{S.~K.} \bibnamefont{Adhikari}},
  \bibnamefont{and}
  \bibinfo{author}{\bibfnamefont{A.}~\bibnamefont{Bala\v{z}}},
  \bibinfo{journal}{Computer Physics Communications}
  \textbf{\bibinfo{volume}{209}}, \bibinfo{pages}{190 } (\bibinfo{year}{2016}),
  ISSN \bibinfo{issn}{0010-4655},
  \urlprefix\url{http://www.sciencedirect.com/science/article/pii/S0010465516302272}.

\bibitem[{\citenamefont{{L. Young-S., D. Vudragovi\'c, P. Muruganandam, S. K.
  Adhikari, and A. Bala\v{z}}}(2016)}]{Young:2016}
\bibinfo{author}{\bibnamefont{{L. Young-S., D. Vudragovi\'c, P. Muruganandam,
  S. K. Adhikari, and A. Bala\v{z}}}}, \bibinfo{journal}{Computer Physics
  Communications} \textbf{\bibinfo{volume}{204}}, \bibinfo{pages}{209}
  (\bibinfo{year}{2016}),
  \urlprefix\url{https://www.sciencedirect.com/science/article/pii/S001046551630073X}.

\bibitem[{\citenamefont{{B. Satari\ifmmode \acute{c}\else \'c{}\fi{}, V.
  Slavni\ifmmode \acute{c}\else \'c{}\fi{}, A. Belic, A. Bala\ifmmode
  \check{z}\else \v{z}\fi{}, P. Muruganandam, and S. K.
  Adhikari}}(2016)}]{Sataric:2016}
\bibinfo{author}{\bibnamefont{{B. Satari\ifmmode \acute{c}\else \'c{}\fi{}, V.
  Slavni\ifmmode \acute{c}\else \'c{}\fi{}, A. Belic, A. Bala\ifmmode
  \check{z}\else \v{z}\fi{}, P. Muruganandam, and S. K. Adhikari}}},
  \bibinfo{journal}{Computer Physics Communications}
  \textbf{\bibinfo{volume}{200}}, \bibinfo{pages}{411} (\bibinfo{year}{2016}),
  \urlprefix\url{https://www.sciencedirect.com/science/article/pii/S0010465515004440}.

\bibitem[{\citenamefont{Young-S. et~al.}(2017)\citenamefont{Young-S.,
  Muruganandam, Adhikari, Lon\v{c}ar, Vudragovi\'c, and
  Bala\v{z}}}]{Young:2017}
\bibinfo{author}{\bibfnamefont{L.~E.} \bibnamefont{Young-S.}},
  \bibinfo{author}{\bibfnamefont{P.}~\bibnamefont{Muruganandam}},
  \bibinfo{author}{\bibfnamefont{S.~K.} \bibnamefont{Adhikari}},
  \bibinfo{author}{\bibfnamefont{V.}~\bibnamefont{Lon\v{c}ar}},
  \bibinfo{author}{\bibfnamefont{D.}~\bibnamefont{Vudragovi\'c}},
  \bibnamefont{and}
  \bibinfo{author}{\bibfnamefont{A.}~\bibnamefont{Bala\v{z}}},
  \bibinfo{journal}{Computer Physics Communications}
  \textbf{\bibinfo{volume}{220}}, \bibinfo{pages}{503 } (\bibinfo{year}{2017}),
  ISSN \bibinfo{issn}{0010-4655},
  \urlprefix\url{http://www.sciencedirect.com/science/article/pii/S0010465517302321}.

\bibitem[{\citenamefont{{Kishor Kumar} et~al.}(2019)\citenamefont{{Kishor
  Kumar}, Lončar, Muruganandam, Adhikari, and Balaz}}]{Kumar:2019}
\bibinfo{author}{\bibfnamefont{R.}~\bibnamefont{{Kishor Kumar}}},
  \bibinfo{author}{\bibfnamefont{V.}~\bibnamefont{Lončar}},
  \bibinfo{author}{\bibfnamefont{P.}~\bibnamefont{Muruganandam}},
  \bibinfo{author}{\bibfnamefont{S.~K.} \bibnamefont{Adhikari}},
  \bibnamefont{and} \bibinfo{author}{\bibfnamefont{A.}~\bibnamefont{Balaz}},
  \bibinfo{journal}{Computer Physics Communications}
  \textbf{\bibinfo{volume}{240}}, \bibinfo{pages}{74 } (\bibinfo{year}{2019}),
  ISSN \bibinfo{issn}{0010-4655},
  \urlprefix\url{http://www.sciencedirect.com/science/article/pii/S0010465519300827}.

\bibitem[{\citenamefont{Nikitin and Pitaevskii}(2005)}]{Nikitin:2005}
\bibinfo{author}{\bibfnamefont{E.~E.} \bibnamefont{Nikitin}} \bibnamefont{and}
  \bibinfo{author}{\bibfnamefont{L.~P.} \bibnamefont{Pitaevskii}},
  \bibinfo{journal}{arxiv:cond-mat/050868v41}  (\bibinfo{year}{2005}).

\bibitem[{\citenamefont{{Roger R. Sakhel, Asaad R. Sakhel, Humam B.
  Ghassib}}(2013)}]{Sakhel:2013}
\bibinfo{author}{\bibnamefont{{Roger R. Sakhel, Asaad R. Sakhel, Humam B.
  Ghassib}}}, \bibinfo{journal}{J. Low. Temp. Phys.}
  \textbf{\bibinfo{volume}{173}}, \bibinfo{pages}{177} (\bibinfo{year}{2013}).

\bibitem[{\citenamefont{Sakhel et~al.}(2016)\citenamefont{Sakhel, Sakhel,
  Ghassib, and Bala\v{z}}}]{Sakhel:2016}
\bibinfo{author}{\bibfnamefont{R.~R.} \bibnamefont{Sakhel}},
  \bibinfo{author}{\bibfnamefont{A.~R.} \bibnamefont{Sakhel}},
  \bibinfo{author}{\bibfnamefont{H.~B.} \bibnamefont{Ghassib}},
  \bibnamefont{and}
  \bibinfo{author}{\bibfnamefont{A.}~\bibnamefont{Bala\v{z}}},
  \bibinfo{journal}{The European Physical Journal D}
  \textbf{\bibinfo{volume}{70}}, \bibinfo{pages}{66} (\bibinfo{year}{2016}),
  ISSN \bibinfo{issn}{1434-6079},
  \urlprefix\url{https://doi.org/10.1140/epjd/e2016-60085-2}.

\bibitem[{\citenamefont{Sakhel}(2016)}]{Sakhel:2016b}
\bibinfo{author}{\bibfnamefont{A.~R.} \bibnamefont{Sakhel}},
  \bibinfo{journal}{Physica B: Condensed Matter}
  \textbf{\bibinfo{volume}{493}}, \bibinfo{pages}{72 } (\bibinfo{year}{2016}),
  ISSN \bibinfo{issn}{0921-4526},
  \urlprefix\url{http://www.sciencedirect.com/science/article/pii/S0921452616301442}.

\bibitem[{\citenamefont{Sakhel and Sakhel}(2019)}]{Sakhel:2019}
\bibinfo{author}{\bibfnamefont{R.~R.} \bibnamefont{Sakhel}} \bibnamefont{and}
  \bibinfo{author}{\bibfnamefont{A.~R.} \bibnamefont{Sakhel}},
  \bibinfo{journal}{Journal of Low Temperature Physics}
  \textbf{\bibinfo{volume}{194}}, \bibinfo{pages}{106} (\bibinfo{year}{2019}),
  ISSN \bibinfo{issn}{1573-7357},
  \urlprefix\url{https://doi.org/10.1007/s10909-018-2068-z}.

\bibitem[{\citenamefont{{R. Onofrio, C. Raman, J. M. Vogels, J. R. Abo-Shaeer,
  A. P. Chikkatur, and W. Ketterle}}(2000)}]{Onofrio:2000}
\bibinfo{author}{\bibnamefont{{R. Onofrio, C. Raman, J. M. Vogels, J. R.
  Abo-Shaeer, A. P. Chikkatur, and W. Ketterle}}}, \bibinfo{journal}{Phys. Rev.
  Lett.} \textbf{\bibinfo{volume}{85}}, \bibinfo{pages}{2228}
  (\bibinfo{year}{2000}).

\bibitem[{\citenamefont{Andrews et~al.}(1996)\citenamefont{Andrews, Mewes, van
  Druten, Durfee, Kurn, and Ketterle}}]{Andrews:1996}
\bibinfo{author}{\bibfnamefont{M.~R.} \bibnamefont{Andrews}},
  \bibinfo{author}{\bibfnamefont{M.-O.} \bibnamefont{Mewes}},
  \bibinfo{author}{\bibfnamefont{N.~J.} \bibnamefont{van Druten}},
  \bibinfo{author}{\bibfnamefont{D.~S.} \bibnamefont{Durfee}},
  \bibinfo{author}{\bibfnamefont{D.~M.} \bibnamefont{Kurn}}, \bibnamefont{and}
  \bibinfo{author}{\bibfnamefont{W.}~\bibnamefont{Ketterle}},
  \bibinfo{journal}{Science} \textbf{\bibinfo{volume}{273}},
  \bibinfo{pages}{84} (\bibinfo{year}{1996}).

\bibitem[{\citenamefont{Ketterle et~al.}(1999)\citenamefont{Ketterle, Durfee,
  and Stamper-Kurn}}]{Ketterle:1999}
\bibinfo{author}{\bibfnamefont{W.}~\bibnamefont{Ketterle}},
  \bibinfo{author}{\bibfnamefont{D.~S.} \bibnamefont{Durfee}},
  \bibnamefont{and} \bibinfo{author}{\bibfnamefont{D.~M.}
  \bibnamefont{Stamper-Kurn}}, in \emph{\bibinfo{booktitle}{Proceedings of the
  Enrico Fermi International School of Physics}}, edited by
  \bibinfo{editor}{\bibfnamefont{M.}~\bibnamefont{Inguscio}},
  \bibinfo{editor}{\bibfnamefont{S.}~\bibnamefont{Stringari}},
  \bibnamefont{and} \bibinfo{editor}{\bibfnamefont{C.~E.} \bibnamefont{Wieman}}
  (\bibinfo{publisher}{IOS Press}, \bibinfo{address}{Amsterdam},
  \bibinfo{year}{1999}), vol. \bibinfo{volume}{CXL}, p.~\bibinfo{pages}{67}.

\bibitem[{\citenamefont{Mendon\ifmmode~\mbox{\c{c}}\else \c{c}\fi{}a
  et~al.}(2018)\citenamefont{Mendon\ifmmode~\mbox{\c{c}}\else \c{c}\fi{}a,
  Ter\ifmmode~\mbox{\c{c}}\else \c{c}\fi{}as, and Gammal}}]{Mendonca:2018}
\bibinfo{author}{\bibfnamefont{J.~T.}
  \bibnamefont{Mendon\ifmmode~\mbox{\c{c}}\else \c{c}\fi{}a}},
  \bibinfo{author}{\bibfnamefont{H.}~\bibnamefont{Ter\ifmmode~\mbox{\c{c}}\else
  \c{c}\fi{}as}}, \bibnamefont{and}
  \bibinfo{author}{\bibfnamefont{A.}~\bibnamefont{Gammal}},
  \bibinfo{journal}{Phys. Rev. A} \textbf{\bibinfo{volume}{97}},
  \bibinfo{pages}{063610} (\bibinfo{year}{2018}),
  \urlprefix\url{https://link.aps.org/doi/10.1103/PhysRevA.97.063610}.

\bibitem[{\citenamefont{{Z.-Ch. Wang, H.-W. Yang, and S.
  Yin}}(2002)}]{Wang:2002}
\bibinfo{author}{\bibnamefont{{Z.-Ch. Wang, H.-W. Yang, and S. Yin}}},
  \bibinfo{journal}{Eur. Phys. J. D} \textbf{\bibinfo{volume}{20}},
  \bibinfo{pages}{117} (\bibinfo{year}{2002}).

\bibitem[{\citenamefont{{Eugene Merzbacher}}(1998)}]{Merzbacher:1998}
\bibinfo{author}{\bibnamefont{{Eugene Merzbacher}}},
  \emph{\bibinfo{title}{{Quantum Mechanics}}} (\bibinfo{publisher}{{John Wiley
  and Sons, Inc.}}, \bibinfo{address}{{New York}}, \bibinfo{year}{1998}),
  \bibinfo{edition}{{Third}} ed.

\bibitem[{\citenamefont{Griffiths and Schr\"oter}(2018)}]{Griffiths:2018}
\bibinfo{author}{\bibfnamefont{D.~J.} \bibnamefont{Griffiths}}
  \bibnamefont{and} \bibinfo{author}{\bibfnamefont{D.~F.}
  \bibnamefont{Schr\"oter}}, \emph{\bibinfo{title}{Introduction to Quantum
  Mechanics}} (\bibinfo{publisher}{Cambridge University Press},
  \bibinfo{year}{2018}), \bibinfo{edition}{3rd} ed.

\bibitem[{\citenamefont{{A. Ronveaux, ed.}}(1995)}]{Ronveaux:1995}
\bibinfo{author}{\bibnamefont{{A. Ronveaux, ed.}}},
  \emph{\bibinfo{title}{{Heun's Differential Equations}}}
  (\bibinfo{publisher}{{Oxford}}, \bibinfo{address}{{New York}},
  \bibinfo{year}{1995}), \bibinfo{edition}{{First}} ed.

\bibitem[{\citenamefont{Sinha and Santos}(2007)}]{Sinha:2007}
\bibinfo{author}{\bibfnamefont{S.}~\bibnamefont{Sinha}} \bibnamefont{and}
  \bibinfo{author}{\bibfnamefont{L.}~\bibnamefont{Santos}},
  \bibinfo{journal}{Phys. Rev. Lett.} \textbf{\bibinfo{volume}{99}},
  \bibinfo{pages}{140406} (\bibinfo{year}{2007}),
  \urlprefix\url{https://link.aps.org/doi/10.1103/PhysRevLett.99.140406}.

\bibitem[{\citenamefont{Tsatsos and Lode}(2015)}]{Tsatsos:2015}
\bibinfo{author}{\bibfnamefont{M.~C.} \bibnamefont{Tsatsos}} \bibnamefont{and}
  \bibinfo{author}{\bibfnamefont{A.~U.~J.} \bibnamefont{Lode}},
  \bibinfo{journal}{Journal of Low Temperature Physics}
  \textbf{\bibinfo{volume}{181}}, \bibinfo{pages}{171} (\bibinfo{year}{2015}),
  ISSN \bibinfo{issn}{1573-7357},
  \urlprefix\url{https://doi.org/10.1007/s10909-015-1335-5}.

\end{thebibliography}

\end{document}